\documentclass[sigconf,nonacm]{acmart}
\usepackage{booktabs}
\usepackage{tabularx}
\usepackage{enumitem}
\usepackage{xcolor}
\usepackage{graphicx}
\usepackage{listings}
\usepackage{pdfpages}
\usepackage{multicol}

\setcopyright{none}
\definecolor{surveyaccent}{HTML}{24557A}
\definecolor{surveylight}{HTML}{EEF3F7}
\definecolor{darkgreen}{HTML}{006400}
\definecolor{revisionbrown}{HTML}{8B4513}
\newcommand{\surveysection}[2]{%
  \par\medskip\noindent
  \colorbox{surveylight}{%
    \parbox{\dimexpr\linewidth-2\fboxsep\relax}{%
      \textcolor{surveyaccent}{\bfseries #1}\hfill{\small #2}}}%
  \par\smallskip}
\newcommand{\surveyitem}[2]{%
  \par\medskip\noindent
  \textbf{\textcolor{surveyaccent}{#1}}\quad #2\par}
\newcommand{\responseformat}[1]{%
  \noindent{\small\textit{Response format:} #1}\par}
\newcommand{\ratingscale}[2]{%
  \noindent{\small\textit{Scale anchors:} 1 = #1; 5 = #2.}\par\smallskip}
\newcommand{\taskratingmatrix}{%
  \ratingscale{very poor}{very good}
  \begin{center}
  \small
  \begin{tabularx}{0.88\linewidth}{@{}Xccccc@{}}
    \toprule
    Criterion & 1 & 2 & 3 & 4 & 5 \\
    \midrule
    Output quality & $\circ$ & $\circ$ & $\circ$ & $\circ$ & $\circ$ \\
    Task achievement & $\circ$ & $\circ$ & $\circ$ & $\circ$ & $\circ$ \\
    Reliability & $\circ$ & $\circ$ & $\circ$ & $\circ$ & $\circ$ \\
    Verifiability & $\circ$ & $\circ$ & $\circ$ & $\circ$ & $\circ$ \\
    \bottomrule
  \end{tabularx}
  \end{center}}
\AtBeginDocument{%
  }

\begin{document}

\title{AI Soccer Analyst: Stage-Aware and Verifiable Human--AI Collaboration for {Soccer} Data Analysis}

\author{Calvin Yeung}
\affiliation{\institution{Nagoya University}\city{Nagoya}\state{Aichi}\country{Japan}}
\email{yeung.chikwong@g.sp.m.is.nagoya-u.ac.jp}

\author{Keisuke Fujii}
\affiliation{\institution{Nagoya University}\city{Nagoya}\state{Aichi}\country{Japan}}
\email{fujii@i.nagoya-u.ac.jp}

\renewcommand{\shortauthors}{Yeung and Fujii}

\begin{abstract}
Sports data analysts translate domain questions into
insights by combining computation with sport-specific domain expertise. Large language
models ease programming, but prompt-to-report workflows may obscure
decisions and evidence. We present AI Soccer Analyst, a
mixed-initiative system with revisable stages: Data
Understanding, Problem Definition, Structured Planning, Execution,
Evidence-Grounded Reporting, and Interaction and Refinement. A formative study
with five analysts first informed design goals for automation, verifiability,
human control, and accessibility. Subsequently, a task-based evaluation with 16
participants combined system logs, retained artifacts, ratings, and open
responses; 33 of 48 tasks met the operational completion criteria.
Exploratory tests supported favorable participant perceptions of completed-task
output quality, task achievement, reliability, and verifiability after Holm
correction. Interaction
records showed domain knowledge emerging through
clarification, planning, and refinement. These findings position stage-aware
human--AI collaboration as a practical approach for producing inspectable,
revisable, and verifiable analyses while retaining domain-expert involvement in
consequential decisions.
\end{abstract}



\begin{teaserfigure}
  \centering
  \includegraphics[width=\textwidth]{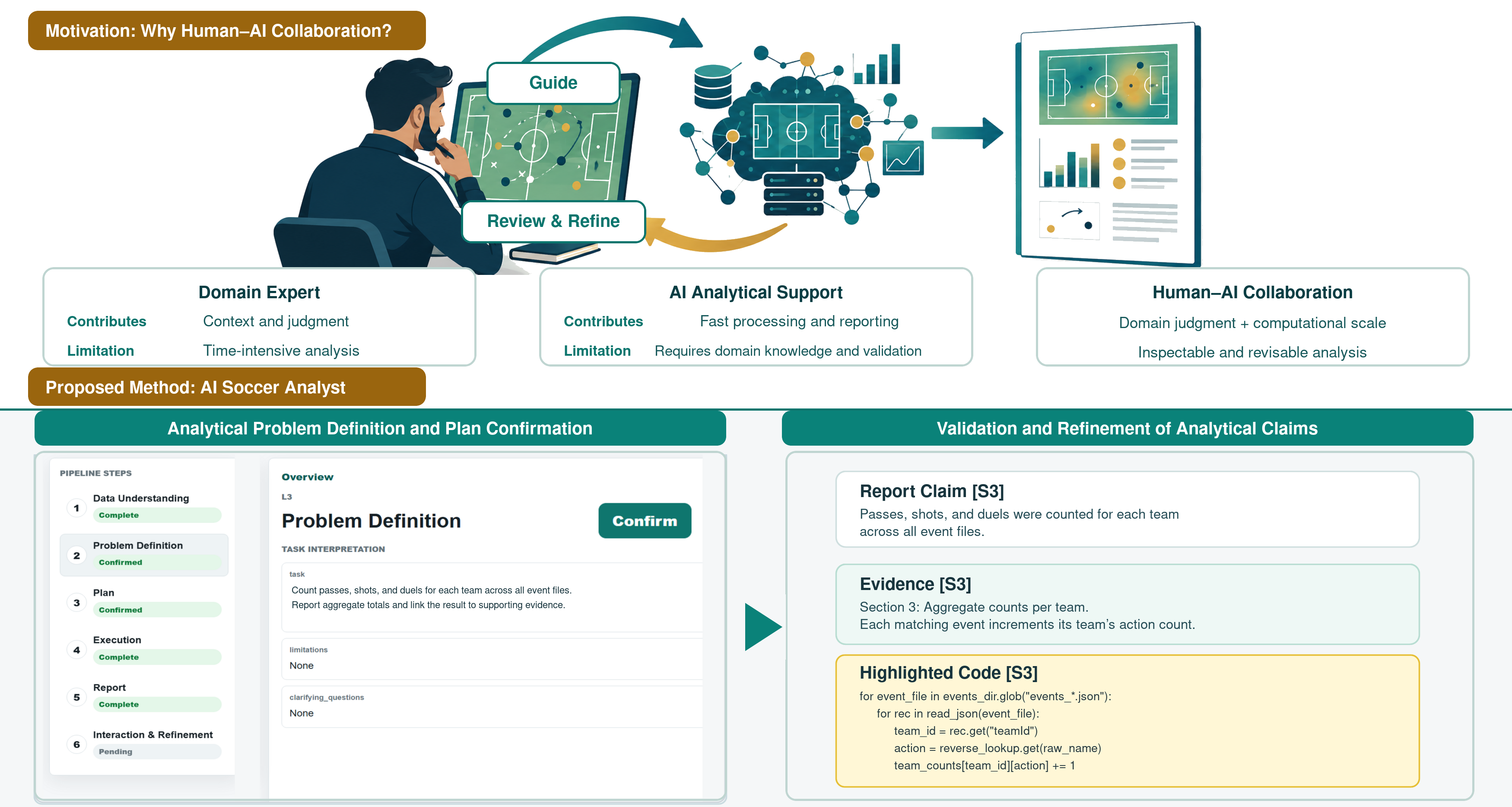}
  \caption{AI Soccer Analyst supports stage-aware human--AI collaboration in
  soccer data analysis. Domain experts contribute context and judgment, while
  AI provides scalable processing and reporting through an iterative
  guidance-and-review loop (top). The interface exposes two key control points
  (bottom): confirming the analytical problem and plan before execution, and
  inspecting linked evidence and relevant code to support claim validation and
  refinement.}
  \Description{The upper section shows a soccer analyst and an AI analysis
  system exchanging guidance and review through a bidirectional loop before
  producing an analytical report. Three boxes summarize the contributions and
  limitations of the domain expert and AI, as well as the outcome of their
  collaboration. The lower section shows the AI Soccer Analyst interface: a
  problem-definition view with a Confirm button on the left and a report claim
  linked to an evidence explanation and code excerpt on the right.}
  \label{fig:teaser}
\end{teaserfigure}


\maketitle

\section{Introduction}

Data increasingly informs decisions in professional practice, but recorded
data does not answer practical questions by itself. Analysts must translate
domain questions into operational definitions, determine what the available
data can support, implement appropriate computations, examine intermediate
results, and communicate evidence in a form that decision makers can use. Data
analysis is therefore an iterative combination of computational work and
situated domain judgment rather than a direct transformation from data to
conclusions \cite{kandel2012enterprise}.

This challenge is particularly visible in competitive sport, where analytical
results must connect recorded performance data to contextual and tactical
decisions. Coaching staff interpret and communicate data while remaining
responsible for its practical consequences \cite{brewer2025coach}, and
analytical work unfolds across interdependent activities rather than isolated
calculations \cite{lee2025crafting}. Soccer analytics supports a broad range of
tasks, including creating summary statistics, assessing player decisions, and
identifying team-level tactical patterns
\cite{yeung2023interpretable,yeung2024onevsone,yeung2024tactics}. At the team
level, comparing how teams attack
requires analysts to decide where a possession begins and ends, identify
recurring sequences of actions, account for the positions of teammates and
opponents, and choose measures that support a fair comparison between teams.
These decisions also depend on what a particular data provider records and how
its schema represents match events.

Large language models (LLMs) offer a promising way to reduce the computational
burden of this work by translating natural-language questions into plans, code,
and reports. However, an answer that sounds convincing may not show how the
system interpreted the question, whether the dataset contains the required
information, or how it translated domain concepts into calculations. An LLM
may assume that an unavailable event is recorded, or produce a
plausible tactical interpretation from an incomplete calculation. Prior
research similarly shows that users of LLM-assisted data analysis can struggle
to communicate sufficient context and verify generated results
\cite{drosos2024rubberduck,gu2024planning}. Supporting soccer analysts
therefore requires more than converting a prompt directly into a report.

Human--AI collaboration provides a basis for dividing this work between the
analyst and the system. Automated support is well suited to
repetitive data inspection, code generation, execution, and artifact
management, whereas analysts are better positioned to define soccer
concepts, select meaningful comparisons, and judge tactical relevance
\cite{horvitz1999mixed,heer2019agency,shneiderman2020hcai}. Yet existing
sports-oriented AI systems have primarily focused on coaching, engagement, or
personalized feedback, while general analytical assistants rarely account for
provider-specific soccer data and domain definitions. Existing provenance and
verification approaches make analytical histories more visible
\cite{ragan2016provenance,gu2024verification}, but they do not fully address
where analyst input should enter an LLM-generated workflow or how a revision
should propagate through its dependent outputs.

This paper presents AI Soccer Analyst, a system that supports human--AI
collaboration in soccer event-data analysis. Its input is a
structured soccer dataset paired with a natural-language analytical question,
and its output is an evidence-grounded report accompanied by inspectable
intermediate artifacts. Automated review checks computational consistency
across these artifacts, while responsibility for judging domain validity
remains with the analyst. Rather than treating the LLM as a
one-step report generator, the system makes key analytical decisions and their
resulting artifacts visible throughout the workflow. Analysts can inspect how
their question was interpreted, revise definitions and plans before assumptions
propagate, and trace reported claims back to execution evidence. Revisions are
directed to the responsible part of the analysis so that affected outputs can
be updated consistently. The central design principle is therefore to combine
computational automation with meaningful opportunities for domain experts to
shape and verify the analysis. Figure~\ref{fig:teaser} illustrates two key
control points: confirming analytical intent before execution and inspecting
the connection between a report claim and its supporting evidence afterward.

The formative findings informed and refined the design goals concerning
efficiency, verifiability, and human analytical control. A subsequent
evaluation with 16 participants found that 33 of 48 analytical tasks
met the operational completion criteria. In exploratory tests, all four
completed-task ratings remained above the neutral midpoint after Holm
correction~\cite{holm1979simple};
seven of eight overall-system ratings met the unadjusted threshold, but only two
remained significant after Holm correction. More importantly, the
interaction records showed that domain knowledge was not provided only in the
initial request; it emerged as participants examined plans and refined results.
Operationally incomplete tasks were concentrated around understanding what the dataset could support,
while open responses emphasized the need for sustained interaction and stronger
verification. Together, these findings show how a structured and revisable
workflow can support useful human--AI analysis while identifying dataset
understanding and active validation as priorities for future systems.

This paper makes the following contributions:
\begin{enumerate}
  \item A formative study of soccer analysts' current practices and
  expectations that identifies requirements for efficient automation, domain
  control, verifiability, accessibility, and practical communication.

  \item AI Soccer Analyst, an implemented workflow for collaboration between
  analysts and AI that externalizes key decisions and artifacts, requests
  domain input when needed, routes revisions to the responsible stage, and
  connects reported claims to execution evidence.

  \item A mixed-method evaluation across analytical tasks of different
  complexity, providing empirical findings on perceived usefulness,
  operational incompletion, the points at which analysts introduce domain
  knowledge, and the interaction and verification support required for
  continued use.
\end{enumerate}

\section{Related Work}

\subsection{Sports Analytics in Practice}

Data analysis involves both computation and interpretation
\cite{kandel2012enterprise}. This is especially important in sport, where the
usefulness of a result depends on the setting in which it is produced. Studies
of collegiate coaching staff show that data use is closely tied to
responsibility for performance and athlete welfare \cite{brewer2025coach}.
Athletes also decide how to use performance data based on their goals,
relationships with coaches, and well-being \cite{brewer2026performance}.
Research on esports coaching likewise shows that analysis is part of wider
coaching and communication practices \cite{lee2025crafting}. Sports analytics
is therefore situated work: a metric gains value only when people can relate it
to their goals and decisions.

SportsHCI research also shows that feedback is more useful when it reflects
expert knowledge. Some systems express coaching principles directly, while
others connect measured performance to the practice context
\cite{weng2025bridging,chen2026badminsense,hirano2026solecoach}. Conversational
coaching agents use a similar idea by adapting support to the user and their
data \cite{jorke2025gptcoach,goldi2025efficient}. Accuracy alone, however, does
not make a system useful. Research on automated ball--strike judgments found
that stakeholders evaluated the system not only by its accuracy but also by how
it changed their roles and decision authority \cite{lee2026ballpark}. Although
automated match decisions differ from analytical assistance, this finding
highlights the need for sports AI to preserve clear opportunities for expert
review.

This need for expert review is especially important in soccer analysis, where
analysts may interpret the same soccer concept in different ways. For
example, one analyst may define an ``attacking play'' as any move toward the
opponent's goal, while another may count only moves that create a scoring
chance. These interpretations produce different measures and may lead to
different conclusions. Structured datasets such as the public Wyscout dataset
provide detailed records, but the data alone cannot determine which
interpretation is appropriate \cite{pappalardo2019wyscout}. Soccer research
has developed advanced methods for representing and analyzing match activity
\cite{yeung2024tactics,yeung2025nmstpp,yeung2025openstarlab}. Yet these methods
offer limited support for a key part of the analyst's work: translating a
soccer question into an analysis whose assumptions, methods, and conclusions
can be clearly understood and evaluated. Our study addresses this gap by
keeping expert interpretation visible throughout the analytical process.

\subsection{LLM-Assisted Data Analysis}

LLMs are increasingly used as conversational interfaces for computational
analysis, translating users' natural-language questions into analytical steps.
Although this can make analysis more accessible, users must still communicate
their intent clearly and determine whether the resulting answer is valid. One
study found that participants using generative AI for data analysis struggled
to provide sufficient context and verify the results
\cite{drosos2024rubberduck}. Consequently, even when the interaction appears
smooth, the system may perform an analysis that does not match the user's
intended question.

This risk begins during planning. Gu et al. separate the decision about what an
analysis should do from the later task of implementing that decision
\cite{gu2024planning}. They found that useful planning support depends on the
current state of the analyst's work. This means that correct code is not enough;
the code must also answer the right question. Domain experts therefore need a
chance to shape the analysis before it runs.

Recent HCI research shows that effective LLM-assisted analysis depends on
preserving user understanding, agency, and opportunities for verification
\cite{shih2024cellsync,guo2024agency,kim2026laps,wu2022aichains,
wu2022promptchainer}. Users can judge and redirect an analysis more effectively
when they can inspect its evolving rationale and outputs rather than only the
final result
\cite{wang2024dataformulator,xie2024waitgpt,vaithilingam2024dynavis}.
Collectively, these findings suggest that user involvement should extend
throughout the analytical process
\cite{amershi2019guidelines,ragan2016provenance,bucinca2021trust}.

Existing sports-focused AI agents tailor coaching and engagement to individual
users and situational contexts
\cite{jorke2025gptcoach,goldi2025efficient,song2025bleacherbot}. However, they
offer limited support for data analysis. By contrast, general-purpose analysis
tools support data analysis but often lack the domain knowledge needed to
interpret soccer-specific concepts correctly. This gap calls for a
soccer-specific workflow that keeps the link between the user's intent and the
computation
visible.

\subsection{Mixed-Initiative Analytical Workflows}

Mixed-initiative interaction is a collaborative approach in which the user and
the AI system can each initiate actions, with control shifting between them as
the task evolves \cite{horvitz1999mixed}. This distribution of control means
that the user's responsibility also depends on where automation is applied
\cite{parasuraman2000automation}. Human-centered AI therefore aims to
combine automation with meaningful human control
\cite{heer2019agency,shneiderman2020hcai}. Users need to understand what the
system is doing and be able to correct important decisions
\cite{amershi2019guidelines}. In data analysis, the right balance may change
during a single task because the AI system provides computational capabilities,
while the analyst contributes domain expertise.

Making the AI system's analytical steps visible can help maintain this balance.
AI Chains and PromptChainer allow users to inspect and change intermediate outputs
\cite{wu2022aichains,wu2022promptchainer}. A study of AI-assisted data analysis
also found that visible assumptions and task steps made the process easier to
guide and check \cite{kazemitabaar2024steering}. These systems provide more
than a sequence of smaller tasks. They give the user a shared view of the work
and show how an early choice affects the final result.

However, each request for user input takes time and attention. Research on
multi-step agents shows that confirmation is useful when it helps users find
errors early, but poorly timed checks can interrupt the task
\cite{zhou2026confirmation}. {AI Soccer Analyst therefore uses
stage-aware intervention: the user confirms the system's understanding and
analytical plan at explicit checkpoints; outside these checkpoints, the system
requests input only when an unresolved decision can materially change the
analysis. Routine work otherwise continues automatically.} Our study
examines whether this stage-aware approach provides control without requiring
constant oversight.

\subsection{Analytical Verifiability and Provenance}

An explanation may help a user understand a result, but it does not prove that
the result is correct. Human-centered XAI therefore argues that explanations
should be designed for the people and practices that use them
\cite{ehsan2020hcxai}. Bansal et al. found that explanations can increase
acceptance without improving human--AI team performance
\cite{bansal2021whole}. The goal should not be to maximize trust, but to help
users judge when the system deserves trust \cite{lee2004trust}. Asking users
to stop and reconsider AI advice can reduce overreliance, although it requires
extra effort \cite{bucinca2021trust}.

Generative AI makes this judgment harder because an incorrect answer can still
sound convincing \cite{tankelevitch2024metacognitive}. In data analysis, users
often need to reconstruct the system's work before they can check the result.
Gu et al. found that analysts used several views of an AI-generated analysis to
understand what the system had done \cite{gu2024verification}. Verification
therefore requires access to the analysis process, not only an explanation of
the final answer.

Analytical provenance addresses this need by recording the history of an
analysis \cite{ragan2016provenance}. In an LLM-assisted workflow, this history
must show how human decisions relate to automated work. We call this
\emph{process verifiability}: users should be able to follow how an analytical
decision becomes reported evidence and revise the process when the connection
is wrong.

Prior research shows that sports analysis depends on its practical context
\cite{brewer2025coach,brewer2026performance,lee2025crafting}. It also shows
that experts need meaningful control over AI-assisted work
\cite{horvitz1999mixed,heer2019agency,amershi2019guidelines,
kazemitabaar2024steering}. Other studies explain how users can inspect an
analysis after it has been produced
\cite{gu2024verification,ragan2016provenance}.
These concerns have rarely been brought together. Our study examines whether a
stage-aware soccer-analysis workflow can support human control and verification
while still providing the efficiency expected from LLM assistance.

\section{Formative Study}
\label{sec:formative-study}

The formative study examined current soccer-analysis practices, workflow
difficulties, and participants' expectations and concerns regarding AI
assistance in soccer analysis. The thematic findings synthesize related
deductive and inductive codes rather than corresponding one-to-one with
individual codes. The complete survey and
follow-up interview instruments are provided in
Appendix~\ref{app:survey}; detailed quantitative results and the qualitative
codebook are provided in Appendix~\ref{app:formative-analysis}.

\subsection{Method}
\label{sec:formative-method}

The formative study employed a survey with selective semi-structured follow-up
interviews. In total, 7 eligible soccer analysts working in collegiate or
professional contexts were contacted; 5 agreed to participate and completed
the survey, and 2 of those 5 also completed follow-up interviews. The
interviews were conducted selectively when participants' open-ended responses
required further clarification or elaboration. The 5 survey participants
were assigned the pseudonymous identifiers E1--E5 according to the
chronological order of their responses; {their characteristics are summarized in
Table~\ref{tab:formative-experience} in
Appendix~\ref{app:formative-analysis}.} The study protocol was approved by the institutional ethics review board and
conducted in accordance with institutional guidelines for research involving
human participants. All participants provided written informed consent before
taking part.

The protocol was guided by the initial assumption that AI assistance in soccer
analysis should address efficiency, verifiability, and
appropriately calibrated trust. The pre-interview survey therefore served two
purposes. First, it examined how participants understood and prioritized these
anticipated requirements. Second, it established a structured overview of
their current practices, including existing workflows, recurring difficulties,
validation practices, and uses of manual, automated, and AI-assisted methods.
For the 2 interviewees, the follow-up questions clarified their survey
responses and probed the reasons and practical context behind them. The
qualitative survey responses from all 5 participants and the follow-up
interview data from 2 participants were analyzed together as a single
qualitative corpus.

Given the small formative sample, closed-ended responses were
summarized descriptively to contextualize the qualitative findings rather than
treated as inferential evidence. {The qualitative analysis used a
combined deductive--inductive approach \cite{braun2006thematic}. Efficiency,
verifiability, and trust were specified a priori as theory-driven sensitizing
concepts and formed the initial deductive categories. Open-ended survey and
follow-up responses were segmented into meaningful units, each of which could
receive multiple codes. A second pass formed and consolidated inductive codes
for content not captured by the initial categories. Accordingly, the formative
findings informed and refined the design goals rather than independently
establishing them.} The analysis was exploratory and conducted by a single
coder. The full codebook and coding results are reported in
Appendix~\ref{app:formative-analysis}.

\subsection{Thematic Findings}
\label{sec:formative-findings}

The analysis yielded five themes concerning where AI assistance may fit within
soccer-analysis practice and the conditions necessary for its responsible use.
These themes address workflow efficiency, verifiability and trust, boundaries
of automation, practical communication, and secure and accessible use.

\paragraph{{Efficiency concerns spanned the analytical workflow.}}
The specific bottlenecks varied by work setting. Participants described substantial effort
in data preparation, report production, and result verification.
E2 described how manual data collection constrained both time and
analytical scope: ``Aggregating data such as phases of play and event-location
coordinates took more than a day and a half. Because the work was slow, the
team asked me to limit data collection to attacking play only.'' This account
also identified broader data-access and infrastructure constraints.
Although these constraints provide important context, they largely fall
outside what an analysis-support system could address. E5 described a highly
automated workflow that still required substantial human effort for
interpretation and verification. Together, these accounts highlighted
opportunities for AI assistance to reduce effort in data preparation and
subsequent stages of the analytical workflow once source data became available.

\paragraph{Verifiability supports calibrated trust.}
Respondents described verification practices combining process inspection,
source-level checking, and domain validation. E2 reported that current data
limitations prevented meaningful validity checking and characterized the
available checking as qualitative rather than objective. Trust was associated
with interpretability, alignment with soccer knowledge, and communication
of uncertainty. Together, these responses suggest that participants considered
access to the underlying data and analytical process necessary for judging
whether an output could be trusted; an explanation alone was insufficient.

\paragraph{{Automation boundaries reflected human analytical
responsibility.}}
{Respondents distinguished routine, time-consuming work from
consequential analytical decisions that they expected to remain under human
control. E5 warned that automation could undermine expertise: ``The
ability to think could be taken away, verification could consume time, and
junior members could lose opportunities to gain experience.''
Together, these responses framed acceptable automation as support for human
judgment without eroding opportunities to develop expertise.}

\paragraph{{Practical communication shaped expectations for
analytical outputs.}}
Respondents emphasized that outputs must fit downstream decision and
communication needs. E1 identified manual report creation and repeated
adjustment of graph design and slide layout as a source of difficulty. E5
noted that output volume could exceed an audience's ability to absorb it.
{Together, these accounts associated useful reporting with
prioritization, fit with existing staff workflows, and opportunities for
focused clarification or revision.}

\paragraph{{Security and accessibility were deployment
concerns.}}
Information leakage and data security were explicit concerns. Participants
{raised protection across data access, model use, storage, and
reporting. Beyond security, they wanted AI-assisted analytical outputs to be
understandable and usable by stakeholders without technical or data-analysis
expertise, such as coaches and players, without continual support from a
specialist analyst.} E3 described the desired outcome as ``an
environment in which anyone can perform analysis independently---where
questions such as `I want that data' or `What is happening with this data?'
can be directed to an LLM rather than a data scientist or data analyst,
allowing people to resolve them independently.''
Together, these responses highlight the importance of protecting information
throughout the analytical workflow while making its outputs accessible to
stakeholders with different levels of technical expertise.

\section{Design Goals}
\label{sec:design-goals}

{The formative findings informed and refined three design goals
that connect the reported needs to the implemented architecture and workflow.
Table~\ref{tab:findings-to-design-goals} summarizes their relationship to the
design goals and {cross-cutting data-exposure and containment requirement}.}

\begin{table*}[t]

  \centering
  \small
  \caption{Relationship between formative findings, design responses, and main
  system implications.}
  \label{tab:findings-to-design-goals}
  \begin{tabularx}{\textwidth}{@{}p{0.22\textwidth}p{0.14\textwidth}X@{}}
    \toprule
    Formative finding & Design response & Main system implication \\
    \midrule
    Workflow-wide efficiency & DG1 & The six-stage workflow automates data
    profiling, plan generation, code execution, review, and reporting from
    available data. \\
    Verifiability and calibrated trust & DG2 & Persistent artifacts link the
    problem definition, confirmed plan, code, execution records, outputs, and
    report claims. \\
    Automation boundaries & DG2 & Confirmation gates precede consequential
    stages, while feedback is routed to the earliest responsible stage for
    regeneration. \\
    Practical communication and self-service use & DG3 & Human-readable,
    evidence-linked reports prioritize supported findings, while the web
    interface supports focused follow-up without requiring code editing. \\
    Data-security concerns & Cross-cutting data-exposure and
    containment requirement & A local model supports data-local
    processing, backend mediation constrains resource access, and an isolated
    coder worker limits some effects of generated code across the workflow. \\
    \bottomrule
  \end{tabularx}
\end{table*}

\phantomsection\label{dg:efficient-data-grounded}
\paragraph{\textbf{DG1: Enable efficient, data-grounded automation.}}
Expectations for greater speed and less manual work extended beyond
computation to preparation, verification, reporting, and revision. External
data-access constraints may
limit the achievable benefit, but the system's primary contribution is to
reduce effort within the analytical stages it supports. DG1 therefore
emphasizes automation across the supported workflow while grounding each
analysis in the data already available.

\phantomsection\label{dg:verifiable-human-control}
\paragraph{\textbf{DG2: Preserve verifiability and human analytical control
{across workflow stages}.}}
Trust and appropriate reliance were associated with transparency into the
analytical process and continued human oversight of domain-specific decisions.
{DG2 therefore emphasizes cross-stage inspectability and
revisability, traceable reported claims, and selective human oversight of
unresolved decisions that could materially change the analysis.}

\phantomsection\label{dg:accessible-actionable-use}
\paragraph{{\textbf{DG3: Support accessible and actionable use
across stakeholders.}}}
{Participants emphasized that analytical outputs needed to be
understandable, appropriately prioritized, and usable by stakeholders with
different levels of technical expertise. DG3 therefore focuses on clear
reporting, accessible interaction, and self-service use without continual
support from specialist analysts.}

\phantomsection\label{req:cross-cutting-security}
\paragraph{{\textbf{Cross-cutting requirement: Reduce
unnecessary data exposure and constrain untrusted operations.}}}
{Data-exposure and containment concerns apply across the design
goals and workflow stages rather than to a single interaction objective. This
yields a cross-cutting requirement to reduce unnecessary data exposure and
limit the effects of untrusted operations.}

\section{System Overview}

\begin{figure*}[t]
  \centering
  \includegraphics[width=\textwidth]{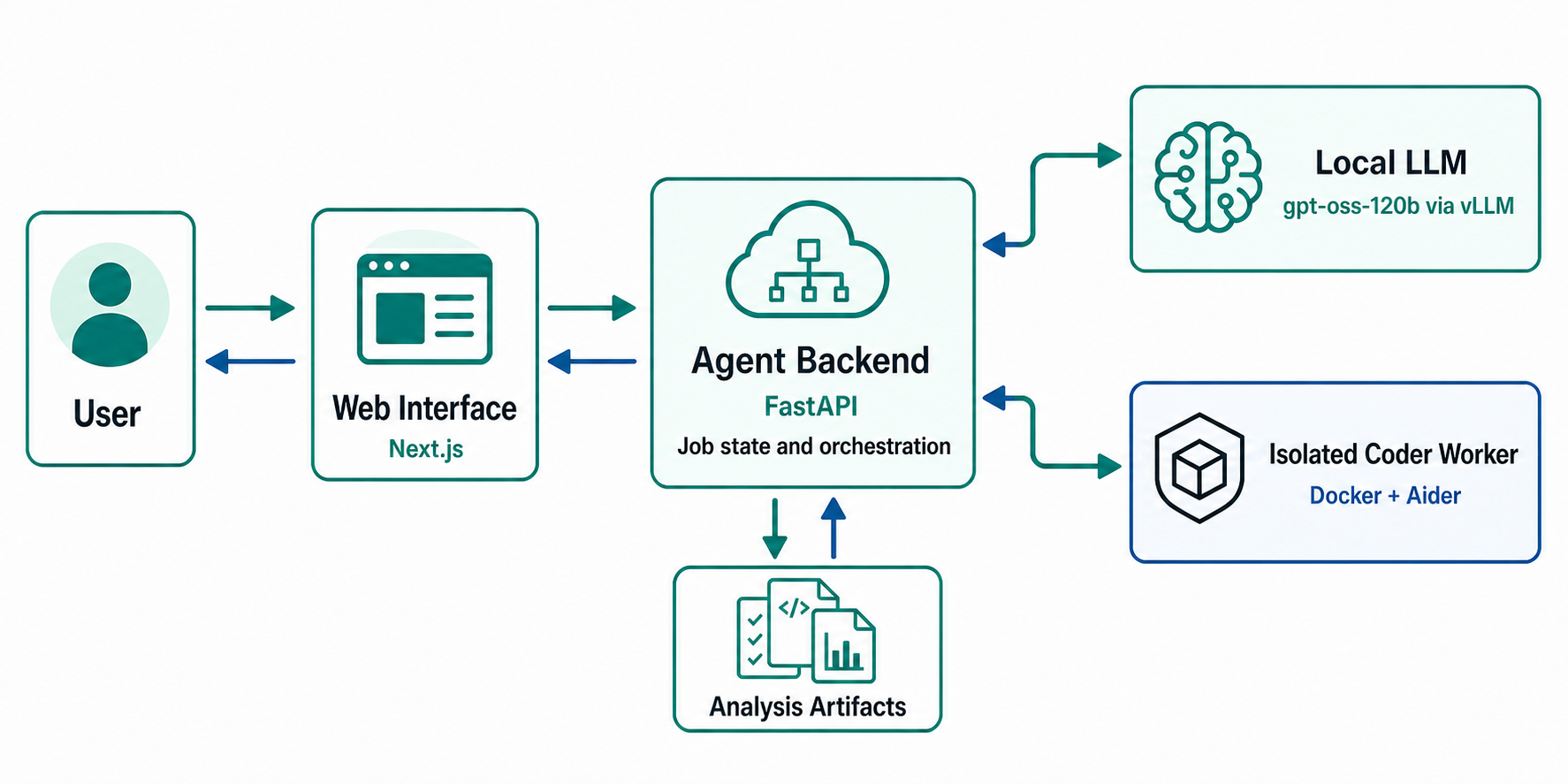}
  \caption{System architecture of AI Soccer Analyst. The web interface communicates with the agent backend, which coordinates the local LLM, isolated coder worker, and analysis artifacts. These artifacts record the problem definition, analytical plan, execution records, generated outputs, review decisions, report, and interaction history across workflow stages.}
  \Description{A user exchanges information with a Next.js web interface, which communicates with a FastAPI agent backend. The backend exchanges requests and results with a locally deployed language model and an isolated coder worker. It reads and writes analysis artifacts containing intermediate and final workflow records so an analysis can retain state across stages.}
  \label{fig:system-architecture}
\end{figure*}

Building on the design goals presented in
Section~\ref{sec:design-goals}, AI Soccer Analyst is a web application that
enables users to create, inspect, and refine soccer data analyses through
natural language. Its web interface supports question submission, artifact
inspection, and feedback, while the backend preserves job and interaction state
and coordinates specialized agents for problem definition, planning, result
review, reporting, and refinement. A locally deployed LLM supports these agents'
reasoning and generation, and an isolated coder worker hosts a coding agent that
generates, executes, and revises analysis scripts before returning the resulting
artifacts to the backend. {Together, these components automate
bounded stages (DG1), preserve inspectable artifacts and user control (DG2),
provide accessible and actionable interaction (DG3), and {support
the cross-cutting data-exposure and containment requirement}.}
Figure~\ref{fig:system-architecture} summarizes the system components and their primary exchanges.

\subsection{Web Interface}

The web interface is the user's point of contact with the system. Implemented as a Next.js application \cite{vercel2026nextjs}, it consists of a dashboard and stage-specific job pages. Each stage produces a persistent, inspectable output (such as a dataset summary, problem definition, analytical plan, execution record, or report), which we refer to as an \emph{artifact}. These artifacts serve as inputs to subsequent stages. A job represents a persistent analysis workspace that connects a user's request and target dataset with the progress, artifacts, and interaction history of the resulting analysis. The dashboard allows users to create a job by selecting a target dataset and providing a job name. It also lists existing jobs together with their current status and analytical stage.

After starting a job, the user moves through stage-specific pages corresponding to the analytical workflow described in Section~\ref{sec:analytical-workflow}. A persistent stage navigator communicates progress and allows the user to revisit the artifacts and interactions associated with each stage. Each page presents the information and interaction controls relevant to its current stage. While an analysis is running, the interface polls the backend for updated state and displays newly produced artifacts when they become available. In the final report, interactive evidence identifiers make the
connection from claims to supporting results or code directly accessible,
combining DG2's inspectability with DG3's emphasis on usable outputs.
Representative interface views are provided in Appendix~\ref{app:ui}.

\subsection{Agent-Based Analysis Backend}

The agent-based analysis backend is implemented with FastAPI \cite{fastapi2026docs} and has three primary responsibilities: controlling access to system resources, managing analysis jobs, and coordinating the agents that perform the analysis. First, the backend authenticates users, enforces account and job
ownership, and controls access to datasets and generated artifacts, providing
{the resource protection required across the workflow.} The web interface accesses the LLM, coder worker, and analysis resources only through the backend. Second, the backend manages the lifecycle and state of each analysis job. This persistent state allows an analysis to progress efficiently
across stages (DG1) while retaining the intermediate results and interactions
needed for inspection, revision, and resumption (DG2).

Finally, the backend coordinates the specialized agents that perform each stage of the analysis described in Section~\ref{sec:analytical-workflow}. For each stage, it selects the appropriate agent and provides the relevant context, including the user's request, target-dataset information, outputs from previous stages, and subsequent feedback. Reasoning and generation requests are routed to the locally deployed LLM, while code-generation and execution tasks are delegated to the isolated coder worker. The returned outputs are recorded as job artifacts and used to update the job's current state. {The backend pauses at defined confirmation checkpoints and, outside them, only when an unresolved decision can materially change the analysis; otherwise, progression remains automatic.} Refinement requests are routed to the responsible stage so that affected downstream artifacts can be regenerated.

\subsection{Local LLM Deployment}

AI Soccer Analyst uses OpenAI's \texttt{gpt-oss-120b}, an open-weight
mixture-of-experts model designed for reasoning, structured generation, tool
use, and agentic workflows \cite{openai2025gptoss}. We selected the 120B model
to balance reasoning capacity with local deployability and used it as a fixed
baseline across the planning, coding, review, and reporting stages. Its
architecture and benchmark performance are documented by OpenAI
\cite{openai2025gptoss}; we cite these results only to characterize the model,
not as evidence for the interaction and verification mechanisms evaluated in
this work. Comparing alternative model sizes and families is outside the scope
of the paper.

The model is hosted locally using vLLM \cite{kwon2023efficient}.
{Using a local deployment rather than an external API supports
data-local processing.} This keeps analytical data, prompts, and intermediate
workflow context within the controlled deployment environment and avoids
transmitting them to an external model provider. Local deployment also allows
the same model checkpoint and inference configuration to be used consistently
across workflow stages and evaluation conditions.

\subsection{Isolated Coder Worker}
\label{sec:isolated-coder-worker}

The coder worker isolates code generation and execution from the analysis
backend because generated scripts may fail, perform unintended file
operations, consume excessive resources, or access sensitive backend
configuration. AI Soccer Analyst therefore runs the CoderAgent and its
execution environment in a separate Docker container
\cite{docker2026overview}. The container limits access to backend resources,
while time, concurrency, and output-size limits bound automated execution,
{reducing exposure of backend resources and containing some effects
of generated code.} The backend remains responsible for job coordination and
analysis state.

The CoderAgent uses Aider \cite{aider} to generate or revise scripts in a
job-specific working directory. For each request, the backend supplies a
bounded bundle containing the confirmed analytical plan, selected context
files, and execution metadata. Aider applies revisions directly to the working
script, allowing focused changes across execution--review rounds without
regenerating it from scratch. After each run, the worker returns the execution
status, logs, generated code, and output files as analysis artifacts. These
records support subsequent review and retain the evidence required for DG2's
inspection and revision.

\section{Analytical Workflow}
\label{sec:analytical-workflow}

\begin{figure*}[t]
  \centering
  \includegraphics[width=\textwidth]{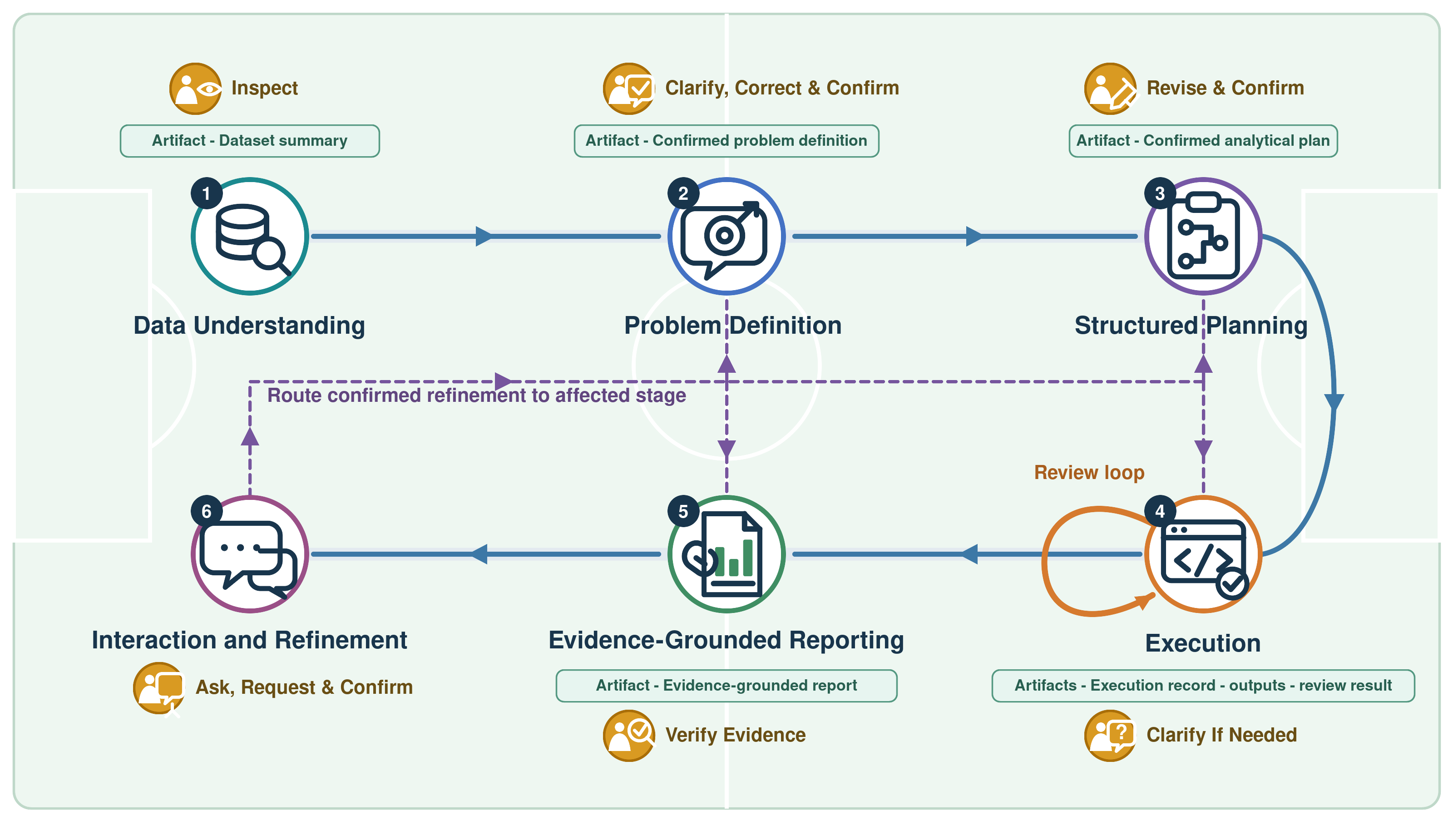}
  \caption{Illustrated six-stage analytical workflow on a soccer-field background. Solid blue arrows show forward progression, the orange arrow shows the internal review loop at Execution, and dashed purple arrows route confirmed refinements from Interaction and Refinement to affected stages. Green pills identify the main artifacts, while orange callouts identify user interactions.}
  \Description{A two-row path connects six illustrated stages. Data Understanding, Problem Definition, and Structured Planning occupy the upper row; Execution, Evidence-Grounded Reporting, and Interaction and Refinement occupy the lower row in reverse reading direction. Blue midpoint arrows show progression from stage 1 through stage 6. The first five stages display green artifact pills for the dataset summary, confirmed problem definition, confirmed analytical plan, execution record and review result, and evidence-grounded report. Orange illustrated callouts show the user actions Inspect; Clarify, Correct and Confirm; Revise and Confirm; Clarify If Needed; Verify Evidence; and Ask, Request and Confirm. An orange looping arrow returns to Execution for review. A dashed purple route rises from Interaction and Refinement, runs across the middle, and branches with direction arrows to Problem Definition, Structured Planning, Execution, and Evidence-Grounded Reporting.}
  \label{fig:analytical-workflow}
\end{figure*}

For the user, the workflow begins with dataset selection and ends with an
evidence-grounded report that they can question or refine. Its six stages are
Data Understanding, Problem Definition, Structured Planning, Execution,
Evidence-Grounded Reporting, and Interaction and Refinement. Execution Review
operates as an internal substage of Execution. Figure~\ref{fig:analytical-workflow}
illustrates the six-stage progression, the main artifacts and user interactions
across the stages, the internal review loop within Execution, and the routes for
confirmed refinements to affected stages. Each stage produces a persistent,
inspectable output, referred to as an \emph{artifact}.
The user confirms the problem and plan at defined checkpoints and
inspects the report and its evidence (DG2). The user otherwise provides input
only when an unresolved decision can materially change the analysis, while
routine workflow steps advance automatically (DG1).
These artifacts support traceability and refinement (DG2) while making the process
accessible to users with different technical capabilities (DG3).

Four agents coordinate the workflow. The AnalystAgent defines the problem,
creates the plan, and produces the report. The CoderAgent implements and
executes the confirmed plan, while the ReviewAgent checks the code and results.
After reporting, the InteractionRefinementAgent answers follow-up questions and
routes requested changes to the responsible stage.
Table~\ref{tab:agent-responsibilities} summarizes each agent's
workflow stages, responsibilities, and main artifacts.
Appendix~\ref{app:agent-prompts} provides shortened agent prompts.

\begin{table*}[t]
\caption{Responsibilities of the four agents in the analytical workflow.}
\label{tab:agent-responsibilities}
\begin{tabularx}{\textwidth}{@{}p{0.16\textwidth}p{0.22\textwidth}X p{0.25\textwidth}@{}}
\toprule
Agent & Workflow stage(s) & Responsibilities & Main artifacts \\
\midrule
AnalystAgent & Problem Definition, Structured Planning, and Evidence-Grounded Reporting & Interprets the user's question, prepares the analytical plan, and synthesizes the final report from reviewed evidence. & Confirmed problem definition, confirmed analytical plan, and evidence-grounded report. \\
CoderAgent & Execution (Analysis Execution) & Uses Aider \cite{aider} to generate or revise the analysis script, which the isolated worker executes to produce the planned outputs. & Execution record and outputs. \\
ReviewAgent & Execution (Execution Review) & Uses a separate review context to compare the confirmed plan, generated code, execution record, and outputs and then accepts the execution, requests revision, or asks for clarification. & Review result. \\
Interaction-\allowbreak RefinementAgent & Interaction and Refinement & Answers questions from existing artifacts, requests clarification when needed, and routes confirmed refinements to the earliest responsible stage. & --- \\
\bottomrule
\end{tabularx}
\end{table*}

\subsection{Data Understanding}

The user begins by selecting the target dataset. The system
automatically profiles its source files, record structures, fields, data types,
missing values, examples, basic statistics, and possible identifier and time
fields (DG1). The system checks what data exists and its structure, but the
user must judge whether the data is sufficient and appropriate for answering
their specific analysis problem (DG2).
The stage ultimately produces a dataset summary stored as JSON for downstream
automation and rendered as Markdown for user inspection.

The summary provides an overview, while task-relevant file details remain
available on demand during planning and review. By summarizing rather than
loading all records, the workflow can accommodate datasets of different sizes
while limiting context use. It is retained as an input to
Structured Planning; Problem Definition focuses only on the user's
natural-language request.

\subsection{Problem Definition}

The user provides a natural-language analytical question and any
constraints or clarifications about the intended entities, scope, comparisons,
and expected result. The AnalystAgent organizes the information stated in the
request and asks for clarification when an ambiguity must be resolved before
planning (DG1). The user then judges whether the resulting definition accurately
represents the intended question, scope, and soccer-domain criteria and
confirms or corrects it (DG2). The stage ultimately produces the confirmed
problem definition.

The interface presents the definition through three fields.
The \emph{Task} field specifies the intended analysis, relevant soccer
entities, scope, comparisons, and expected result. The \emph{Limitations}
field records constraints or preferences provided by the user, while
the \emph{Clarifying questions} field asks the user to resolve parts of the
request that could be interpreted in multiple ways and lead to different
analyses. For example, ``top-performing teams'' does not specify how
performance should be measured or which teams should be included.
While clarifying questions or corrections remain, the workflow
stays in this stage; user confirmation transitions it to Structured Planning.

\subsection{Structured Planning}

Structured Planning begins with the problem definition confirmed
by the user in the preceding stage. During planning, the user may add
constraints or revise the proposed analytical steps. The AnalystAgent combines
that input with the dataset summary and relevant dataset details to translate
the requested soccer concept into a quantitative definition and a sequence of
coding steps. It constructs the plan using available files, fields, values, and
mappings and requests targeted context when essential information is missing.
Constructing the plan from available dataset context supports data-grounded
planning (DG1). The user judges whether the quantitative definitions,
comparisons, and assumptions validly represent the intended soccer concepts
(DG2). The plan presents these choices as user-readable steps (DG3). After the
user confirms them, the stage ultimately produces the confirmed
analytical plan.

For targeted retrieval, the AnalystAgent can call
\nolinkurl{get_file_summary} to inspect a relevant file;
Appendix~\ref{app:planning-retrieval} describes the available functions. Each
result is added to the planning context, and the number of retrieval rounds is
limited. The workflow remains in planning while additional information is
needed or the user is revising the plan, and advances to Analysis Execution
only after user confirmation.

\subsection{Execution}

\subsubsection{\textbf{Analysis Execution}}

The user provides the confirmed analytical plan and, only when an
unresolved decision concerns analytical intent, scope, or preference, any
requested clarification; the user does not need to write or run code. The
CoderAgent automatically generates and executes the analysis script in the
isolated worker (DG1). The worker enforces operational limits and records logs
and outputs that support downstream traceability (DG2).
Analysis Execution itself does not validate these artifacts; that assessment
occurs in Execution Review.
The substage ultimately produces the execution record and analytical outputs
for review.

Behind the interface, the CoderAgent uses Aider in the isolated
worker (Section~\ref{sec:isolated-coder-worker}) to generate or revise an
analysis script. The execution record retains the generated code, logs, and
output summary for the ReviewAgent. The final user-facing report is synthesized
separately by the AnalystAgent during Evidence-Grounded Reporting. A completed
run advances to Execution Review.

\subsubsection{\textbf{Execution Review}}

The user's confirmed plan supplies the review criteria, and the
user provides clarification only when a question concerns analytical intent,
scope, or preference. Using a new context independent of the CoderAgent, the
ReviewAgent compares the confirmed plan, generated code, execution record, and
outputs to assess plan adherence and computational consistency (DG2). The
substage ultimately produces a review result recording acceptance, revision
feedback, or a clarification request.

Most review and revision rounds occur automatically (DG1).
Acceptance advances to Evidence-Grounded Reporting, while revision feedback
returns the workflow to Analysis Execution. Dataset questions are resolved by
retrieving the required details, whereas the user is involved when
clarification concerns analytical intent, scope, or preference. The configured
round limit stops automated progression.

\subsection{Evidence-Grounded Reporting}

Evidence-Grounded Reporting receives the reviewed outputs and
execution records from the preceding stage, together with the cleaned code and
confirmed problem definition. The AnalystAgent uses these artifacts to generate
a report organized around the user's problem (DG1). It grounds report claims in
the reviewed outputs and execution records and connects each claim to its
supporting evidence through identifiers (DG2). The cleaned code is provided as
an explanatory view of the computation, not as evidence that verifies a claim.
The stage outputs the evidence-grounded report with its supporting tables,
figures, and claim-to-evidence associations in a user-readable form (DG3). The
user then checks the validity of the report's interpretations and conclusions
(DG2).
Appendix~\ref{app:l1-code} provides implementation details and an
example.

The AnalystAgent may improve how supported results are organized and explained,
but it cannot fix incorrect calculations.
{A status of ``Complete'' means that the automated workflow
has reached its terminal state; it does not indicate expert validation of
analytical correctness.}
Changes to the analytical definition or computation must
return to the responsible upstream stage. Therefore, the workflow transitions to
Interaction and Refinement, where the user can ask follow-up questions or
request corrections.

\subsection{Interaction and Refinement}

After reading the report, the user either asks a follow-up
question or requests a change. The InteractionRefinementAgent answers a
follow-up question from existing artifacts or, if a refinement is required,
routes it to the earliest responsible stage and identifies only the affected
downstream artifacts (DG1). Before refinement begins, the user confirms that the
proposed target and scope match the intended change and remains responsible
for judging the domain validity of the revised analysis (DG2). The stage
ultimately produces either an artifact-grounded answer to the question (DG3) or
an updated set of affected artifacts and their visible interaction history
(DG2); it introduces no separate main artifact.

When an artifact is revised or regenerated, the new version
replaces the previous version. Consequently, only the current version remains
inspectable. The scope of
regeneration depends on the requested change: a wording change updates only the
report, a change to a metric definition returns to Structured Planning, and a
change to the teams or time period being compared returns to Problem
Definition.

\section{User Experiment}
\label{sec:evaluation-method}

A task-based evaluation of AI Soccer Analyst was conducted to examine
operational completion, perceived output quality, and the evidence participants
used to judge an analysis. This was a single-system evaluation rather than a
controlled comparison with a baseline. Each retained participant attempted one analytical task at each of three levels of increasing analytical difficulty: L1 (descriptive retrieval), L2 (comparative interpretation), and L3 (tactical synthesis). The evaluation combined
system logs and retained artifacts with task-level ratings, an overall-system
survey, and open responses.

\subsection{Dataset}

All analytical tasks used the public Wyscout 2017 soccer-event dataset
\cite{pappalardo2019wyscout}. As reported in the dataset paper, the dataset
contains over 3 million events from over 1.9 thousand matches involving over 4 thousand players across
seven competitions: the top divisions of England, France, Germany, Italy, and
Spain, plus UEFA Euro 2016 and the 2018 FIFA World Cup. The accompanying files
provide match, team, player, competition, and coaching metadata.

Event data is a chronological log of discrete match actions. For example, one
record may describe a pass by a particular player and team at a given match
time and pitch location, while a later record may describe the resulting shot.
Records include event and subevent labels, identifiers for the match, team, and
player, match period and time, event tags, and locations when available. These
fields support filtering, aggregation, sequence analysis, and spatial
summaries. However, these analyses are limited to actions recorded in the event log. For example, the dataset cannot be used to analyze off-the-ball player
  movement or changes in team formation because this information is not recorded.

\subsection{Three Levels of Analytical Tasks}

The tasks were divided into three levels to represent distinct stages of
analytical work rather than treating every task as simply easy or difficult.
L1 tests whether the system can retrieve and summarize explicitly requested
information. L2 introduces the additional need to choose appropriate
comparison conditions and interpret differences in context. L3 tests whether
the system can integrate several event patterns into a coherent tactical
account. Keeping these stages separate makes it possible to evaluate the
system as analytical abstraction, verification effort, and reliance on domain
judgment increase.

The levels also differ in the soccer knowledge needed to formulate and
assess an answer. L1 requires relatively little domain knowledge because its
quantities can be obtained through direct filtering and aggregation. L2
requires moderate domain knowledge to select meaningful comparison groups and
judge whether definitions are applied consistently. L3 requires the most
domain knowledge because several results must be connected to higher-level
soccer concepts and assessed for tactical plausibility. Table~\ref{tab:evaluation-task-levels}
summarizes these distinctions and gives one representative task at each level.
A complete L1 example, including its problem definition, confirmed plan,
report, evidence links, and cleaned code, is provided in
Appendix~\ref{app:l1-example}.

\begin{table*}[t]
  \centering
  \small
  \setlength{\tabcolsep}{5pt}
  \caption{Analytical task levels used in the evaluation.}
  \label{tab:evaluation-task-levels}
  \begin{tabular}{@{}p{0.14\textwidth}p{0.19\textwidth}p{0.27\textwidth}p{0.33\textwidth}@{}}
    \toprule
    Level & Goal & Characteristics & Example \\
    \midrule
    L1---Descriptive Retrieval &
    Retrieve and summarize explicit information from event data. &
    Simple filtering, aggregation, and descriptive statistics with low
    reasoning and verification demands. &
    Summarize the number of passes, shots, and duels for each team. \\
    \addlinespace
    L2---Comparative Interpretation &
    Compare and interpret patterns across multiple analytical conditions. &
    Multi-condition comparison, contextual reasoning, and moderate
    verification effort. &
    Compare passing patterns between Team A and the top-performing teams. \\
    \addlinespace
    L3---Tactical Synthesis &
    Produce higher-level tactical and analytical conclusions from event data. &
    Multi-step reasoning, analytical synthesis, report generation, and
    integration of multiple event patterns. &
    Generate a tactical analysis report describing a team's attacking
    tendencies based on passing sequences, shot creation, and player
    involvement. \\
    \bottomrule
  \end{tabular}
\end{table*}

\subsection{Experiment Method}

Participants attempted the three tasks: L1,
then L2, then L3. Before starting, they received the same written study instructions
and a PowerPoint walkthrough covering the Wyscout dataset, the three task-level
definitions with examples, and the system workflow for creating a job,
confirming the problem definition and plan, and inspecting the report and its
evidence. Participants formulated their own soccer-analysis question at each
level rather than following the example; the level descriptions
and examples in Table~\ref{tab:evaluation-task-levels} guided the intended
scope and complexity. Participants were instructed to spend approximately
45--60 minutes on each task. Jobs could continue asynchronously beyond the session, but a
job that had not produced a final report at the 24-hour cutoff was stopped and
classified as incomplete. If a task failed, was cancelled, or remained
unresolved, participants proceeded to the next level.

After each attempted task, including an incomplete one,
participants recorded the job name and rated output quality, task achievement,
reliability, and verifiability on 5-point Likert scales from \emph{very poor}
to \emph{very good}. For analysis, a task was classified as operationally
complete when the system completed the Evidence-Grounded Reporting stage and
generated a final report; this corresponded to a final top-level job status of
\texttt{complete}. Jobs ending as \texttt{failed} or \texttt{cancelled}, or otherwise
remaining unresolved, were classified as operationally incomplete.
Operational completion indicates successful pipeline completion and report
generation, but does not establish the semantic or statistical correctness of
the analysis.
By this definition, 33 of 48 tasks were operationally complete. All 15
incomplete tasks were rated and retained in the analysis of operationally incomplete tasks, but their
ratings were excluded from the completed-task rating analysis. After the three
task stages, participants completed the overall-system assessment and
open-response questions.

The overall-system assessment contained eight agreement items covering trust, the
ability to judge validity, process and evidence verifiability, efficiency,
reduced manual work, interaction usefulness, and intended future use. In addition, four
open-response questions asked what worked well, what was difficult or
inefficient, which outputs felt untrustworthy, and what additional evidence
would support verification. The complete Survey~2 instrument appears in
Appendix~\ref{app:survey}.

Invitations were sent to university soccer-team analysts, master's or doctoral
students whose research focused on soccer, and individuals who belonged to
both groups. In total, 23 participants were recruited, of whom 16 completed
the experimental procedure and submitted Survey~2. The analysis included all 16 participants and their 48
analytical tasks, with 16 tasks at each level. One participant (P013) had no
operationally completed task but remained in the cohort so that operationally incomplete tasks and
overall-system perceptions were represented. The experiment was approved by
the institutional ethics review board, and all participants provided written
informed consent before taking part.

\section{Results}
\label{sec:evaluation-results}

This section presents five complementary views of the evaluation before
deriving design implications. It first examines participants' ratings of
completed analytical tasks and then identifies why other tasks were operationally incomplete. It next
characterizes human--AI collaboration across the workflow, reports overall
system ratings, and uses the open responses to explain participants' positive
assessments and remaining concerns. Together, these analyses distinguish the
perceived value of completed results from operational barriers and show how
participants contributed domain expertise during the analytical process. Supplementary
information and full statistical results appear in
Appendix~\ref{app:evaluation-analysis}.

\subsection{Ratings for Completed Tasks}

Ratings were summarized descriptively using medians and interquartile
ranges. To explore whether participants perceived completed outputs favorably,
we used exact one-sided Wilcoxon signed-rank tests
\cite{wilcoxon1945individual} to determine whether participant-aggregated ratings
of output quality, task achievement, reliability, and verifiability exceeded the
neutral midpoint of 3 on the 5-point Likert scale (1--5). The analysis
included the 33 operationally completed analytical tasks. These
ratings therefore characterize completed tasks only and do not represent
participants' experiences with operationally incomplete tasks. Ratings were first averaged
within participant for each
outcome because ratings from tasks completed by the same participant are not
independent. This also prevented participants with more completed tasks from
having greater influence on the results. Each of the 15 participants with at least one completed task therefore
contributed a single value per outcome; P013 contributed no completed-task
rating.

\begin{figure}[t]
  \centering
  \includegraphics[width=\columnwidth]{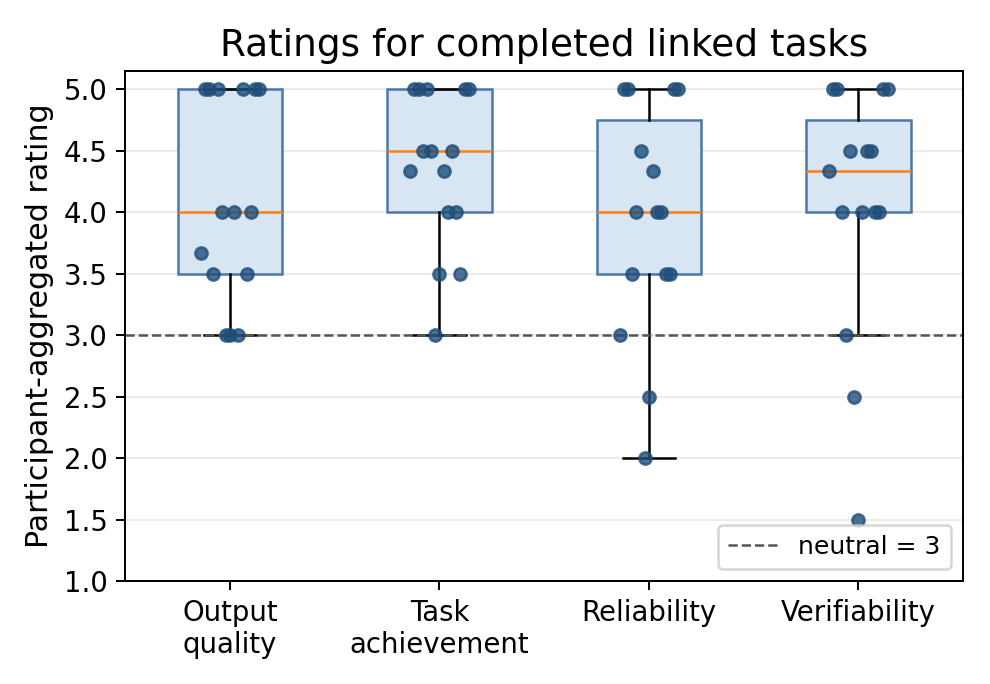}
  \caption{Participant-aggregated ratings for the 33 completed analytical tasks.
  Points represent participants. Each box spans the interquartile range (IQR;
  25th--75th percentiles), its central line marks the median, and its whiskers
  extend to the most extreme values within 1.5 times the IQR. The dashed line
  marks the neutral value of 3.}
  \Description{Four box-and-point plots show participant-aggregated output
  quality, task achievement, reliability, and verifiability ratings. Boxes
  show the interquartile ranges and median lines, with whiskers extending to
  values within 1.5 times the interquartile range. Most values lie above the
  dashed neutral line at 3.}
  \label{fig:evaluation-task-ratings}
\end{figure}

All four participant-aggregated ratings met the unadjusted
$p<.05$ threshold: output quality ($p=.0002$), task achievement ($p=.0001$),
reliability ($p=.0022$), and verifiability ($p=.0022$). All four remained
significant after Holm correction~\cite{holm1979simple} (adjusted
$p=.0002$--$.0044$).
Figure~\ref{fig:evaluation-task-ratings}
shows the distributions of participant-aggregated ratings. Complete descriptive and
participant-level test statistics, together with task-level analyses stratified
by task complexity level (L1--L3), are reported in
Appendix~\ref{app:evaluation-analysis}.
These results indicated that participants who received a completed output
generally evaluated its quality, usefulness for the task, reliability, and
verifiability favorably. The consistently positive ratings suggest
that participants perceived the completed analyses as useful and verifiable,
consistent with the intended goals of accessibility and transparency.

\subsection{Operationally Incomplete Task Analysis}

To identify where the workflow broke down, the 15 incomplete analytical tasks
were classified by their primary cause of incompletion. Classification drew on each
task's final job state, execution review, and artifact status. The causes were
grouped into 5 categories and summarized descriptively, with each task
assigned to a single category.

\begin{figure}[t]
  \centering
  \includegraphics[width=\columnwidth]{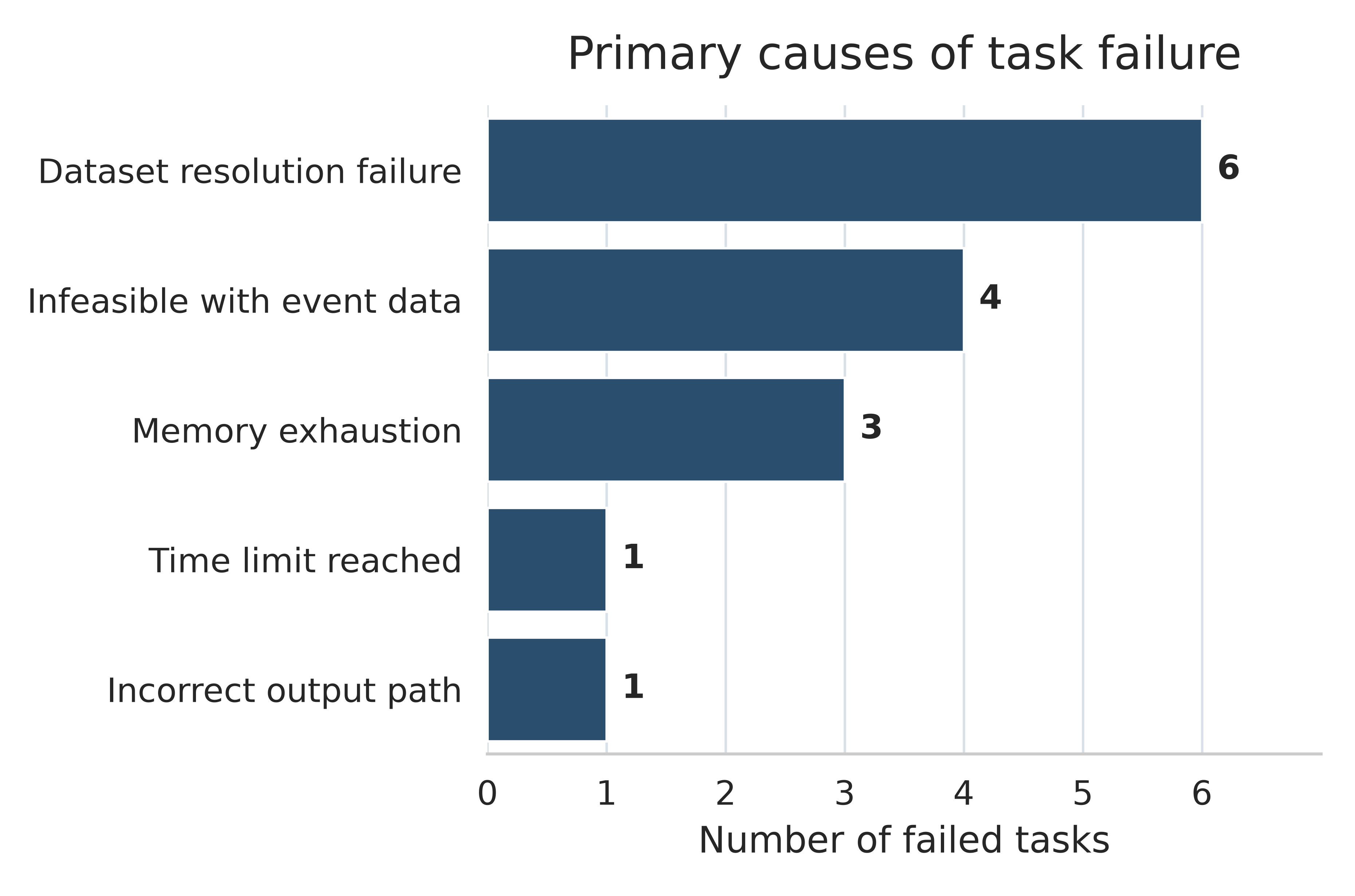}
  \caption{Primary causes assigned to the 15 operationally incomplete analytical tasks.}
  \Description{A horizontal bar chart shows six tasks involving schema, identifier, or data-resolution problems;
  four tasks infeasible with the available event data; 3 tasks
  incomplete because the system ran out of memory; 1 task incomplete because the report was
  saved to the wrong path;
  and 1 cancellation after the experiment time
  limit.}
  \label{fig:evaluation-failures}
\end{figure}

In total, 4 tasks across 2 participants were infeasible with the
available event data because they required tracking or off-ball information.
all three of P013's tasks were operationally incomplete for this reason. A further 6 tasks across 5 participants failed because execution could not resolve
required dataset fields, identifiers, tags, or available records. These tasks
ranged from player rankings and temporal passing comparisons to tactical
reports, showing that schema and data-capability problems occurred across the
three analytical levels. Another 3 tasks stopped because the system ran out of
memory during the analysis, 1 failed because the required report was saved to the wrong path, and 1 was cancelled after reaching the experiment's time limit (24 hours). The distribution of causes of incompletion is shown in
Figure~\ref{fig:evaluation-failures}.

The concentration of operationally incomplete tasks around data resolution highlighted the value of
checking dataset capabilities before execution. The remaining cases
highlighted the need for resource-aware execution, support for resuming
time-limited tasks, and artifact validation before workflow advancement.
Separating infeasible requests from resolvable schema and identifier
failures clarifies which tasks exceeded the information available in the event
data and which failed during execution. These findings qualified
DG1 by illustrating the importance of resolving provider-specific schemas and
dataset capabilities before execution.

\subsection{Human--AI Collaboration Patterns}

To characterize how participants engaged with the stage-aware workflow and how
soccer-domain knowledge entered the analysis, logs from all 48 analytical tasks
were examined for system-requested clarification, feedback on the proposed
plan, and post-result refinement. A domain-knowledge contribution was defined
as participant input that changed the soccer context, analytical scope,
metric or comparison definition, data feasibility, or interpretation.
Presentation-only changes were excluded. Categories could overlap, and all
comparisons were descriptive.

Across the interaction types, system-requested clarification occurred least
often, in 3 of 48 tasks (6.3\%): 2 requests asked participants to identify
the exact team and 1 asked for the exact match. Participants provided plan
feedback in 11 tasks (22.9\%). This feedback addressed scope,
metric definitions, output requirements, and data feasibility; examples
included organizing results by team and defining progressive passes from
player movement rather than an assumed event tag. Participants requested
post-result refinement in 9 tasks (18.8\%), including expanded comparisons,
corrected metric representations, presentation changes, and further
interpretation.

Participants introduced or revised explicit soccer-domain knowledge in 14 of
48 tasks (29.2\%). Its first point of entry was most often Structured Planning
(7/14 tasks), followed by post-result refinement (4/14) and clarification
(3/14).
Participants identified the relevant team or match, refined the analytical
scope, defined soccer measures and comparisons, questioned dataset
feasibility, and reconsidered how results should be interpreted.

Together, these patterns showed that participant involvement extended beyond
answering system questions: analysts more often shaped the plan or revised the
result. Soccer-domain knowledge also entered throughout the workflow rather
than only through the problem definition. Structured Planning was the main stage
at which participant expertise became an executable analytical definition,
while post-result refinement allowed soccer judgment to be applied after
concrete results became visible. These patterns illustrate
mixed-initiative use of the stage-aware workflow: participants expressed domain
expertise through explicit, revisable analytical decisions rather than only
through the initial prompt. These observations were consistent with aspects of
DG2: participants exercised domain control by revising analytical definitions
during planning and after reviewing the results.

\subsection{Overall System Ratings}

Ratings were summarized descriptively using medians and interquartile
ranges. To explore whether participants perceived the system favorably, we used
exact one-sided Wilcoxon signed-rank tests \cite{wilcoxon1945individual} to
determine whether each of the eight overall-system ratings exceeded the neutral
midpoint of 3 on the 5-point Likert scale (1--5). The analysis used a single response per item
from each of the 16 participants.

\begin{figure*}[t]
  \centering
  \includegraphics[width=\textwidth]{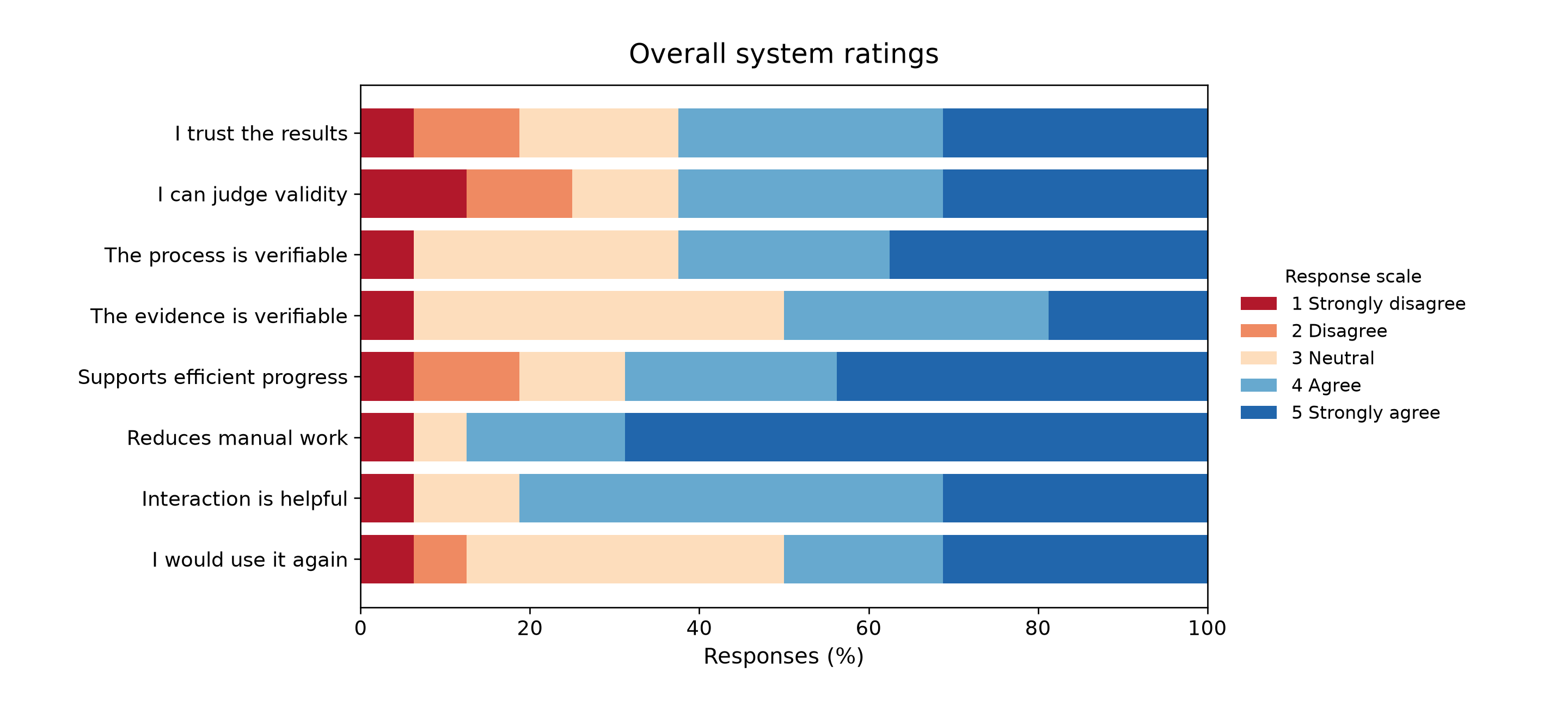}
  \caption{Distributions of the eight overall-system ratings. Ratings range
  from 1 (strongly disagree) to 5 (strongly agree).}
  \Description{Eight horizontal stacked bars show the percentage of responses
  at ratings 1 through 5 for trust, validity judgment, process and evidence
  verification, efficiency, manual-work reduction, interaction, and future
  use. Ratings 4 and 5 form the largest share for most items.}
  \label{fig:evaluation-overall-ratings}
\end{figure*}

Seven of the eight ratings met the unadjusted $p<.05$ threshold
($p=.0006$--$.0449$). After Holm correction~\cite{holm1979simple} across the eight items, only reduced
manual work (adjusted $p=.0049$) and helpful interaction (adjusted $p=.0184$)
remained significant. Responses leaned toward agreement across
all eight items, with the clearest concentration of positive responses for
manual-work reduction and helpful interaction. Responses were more mixed for
validity judgment and evidence verification. Evidence verifiability
met the unadjusted threshold ($p=.0391$) but not the Holm-adjusted threshold
(adjusted $p=.1333$), whereas validity judgment did not reach the unadjusted
threshold ($p=.0873$). Although validity-judgment ratings were
directionally positive (mean $=3.56$, median $=4$), responses varied considerably
(SD $=1.41$, IQR $=2.75$--$5$). This
result does not show that ratings were neutral; rather, the sample provided
insufficient evidence that they consistently exceeded neutral. This suggests
that participants generally found the evidence inspectable, but were less
certain whether the analysis itself was valid. Figure~\ref{fig:evaluation-overall-ratings} shows
the response distributions, and complete item-level statistics appear in
Appendix~\ref{app:evaluation-analysis}.

Participants most clearly perceived practical benefits,
reporting reduced manual work and useful interaction. These
responses were consistent with DG1's efficiency objective and DG3's emphasis on
accessible interaction. Less consistent agreement on validity and evidence
verification identified stronger support for judging whether an analysis
should be trusted as a remaining priority for realizing DG2.

\subsection{Open-Response Themes}

The open responses helped explain why participants evaluated the system as they
did. In total, 9 participants valued analytical efficiency and accessibility because
the system could search large event datasets, automate work, and produce usable
reports without requiring them to write code. Analytical support
and new insights were valued by 6 participants: structured plans, suggested
analyses, and alternative views
helped them decide what to examine and how to interpret a result.
These accounts provided qualitative support for DG1 and DG3 by showing why
manual-work reduction and helpful interaction received high overall ratings.

The improvement-oriented responses formed two broader themes. In total, 11 participants
emphasized interaction continuity and recovery, indicating that an analysis
should remain understandable and revisable when interaction or execution
breaks down. Analytical credibility and verification were emphasized by 11
participants, who sought sufficient evidence to judge analytical definitions, results, and
interpretations before relying on them. The reported numbers indicated how many
participants mentioned each theme. A participant could be included in more
than one theme.
Together with the data-resolution failures reported earlier, these themes
motivated the three design implications presented next. The
open response analysis is detailed in
Appendix~\ref{app:evaluation-analysis}.

\subsection{Design Implications}

The evaluation findings motivated three implications for extending stage-aware
analytical systems. These implications concern how systems understand changing
data sources, sustain human--AI collaboration across analytical tasks, and
support verification beyond tracing how an output was produced.

\paragraph{Maintain provider-specific schema and capability knowledge.}
Soccer data providers differ not only in their schemas,
terminology, and APIs, but also in the analytical capabilities supported by
their data. Future systems should therefore maintain validated,
provider-specific knowledge of available fields, event types, spatial and temporal
granularity, match coverage, and data modalities such as event, tracking, and
off-ball data. This knowledge should be incorporated into dataset understanding
and planning to map analytical concepts to the correct data representations and
assess feasibility before execution. When a request requires information that
is unavailable, such as player movement between events, sprint trajectories,
or off-ball positioning, the system should explain the limitation, identify
the missing data modality, and either request an appropriate data source or
propose a feasible event-data-based alternative. This would reduce schema
resolution failures while preventing inherently unsupported analyses from
proceeding to execution.

\paragraph{Extend refinement into persistent human--AI collaboration.}
The current system routes refinement to the responsible analytical stage and
regenerates its dependent artifacts. Future systems could additionally
preserve alternative analysis branches, allow analysts to compare the effects
of different definitions or plans, and reuse analyst-approved domain
definitions across subsequent tasks. This would transform individual
revisions into persistent knowledge, reduce repeated corrections, and allow
the system to adapt to an analyst's established practices.

\paragraph{Move from traceability to active verification.}
The current system links claims to plans, code, data, and result artifacts,
allowing analysts to trace how an output was produced. Future systems should
also actively evaluate whether those outputs are credible using domain-specific
validation rules, anomaly detection, independent recomputation, and comparison
with authoritative external statistics when available. Conflicting evidence
and unresolved assumptions should be highlighted so analysts can focus their
verification effort on the results most likely to require domain judgment.

Together, these implications extend the stage-aware workflow from supporting
individual analytical tasks toward retaining knowledge across providers,
revisions, and verification activities. They were also grounded directly in
the evaluation: data-resolution failures motivated provider-aware dataset
understanding, observed plan and result revisions motivated persistent human--AI
collaboration, and participants' verification concerns motivated active
validation.

\section{Discussion}

The central contribution of AI Soccer Analyst is not simply the generation of
an analytical answer. It is the organization of analysis as a process in which
system actions remain inspectable and human expertise can change the analysis
at the stage where it is needed. The evaluation provides two complementary
perspectives on this contribution: process artifacts supported inspection and
verification, while clarification, planning feedback, and refinement allowed
participants to introduce domain knowledge as the analysis developed.

\subsection{Revisability as Process-Level Transparency}

AI Soccer Analyst shifts the focus of transparency from explaining a final
answer to exposing the process that produced it. A plausible explanation alone
does not reveal how a question was interpreted, which data were selected, or
how an analytical definition was implemented. Externalizing the problem
definition, plan, execution record, and claim--evidence links makes these
decisions available for inspection and allows a correction to be routed to the
responsible stage. The Holm-adjusted completed-task verifiability
rating and the descriptively favorable overall evidence-verifiability ratings
suggest that participants found this process-level visibility useful. However,
neither the overall evidence-verifiability nor validity-judgment item was
significant after Holm correction, showing that supporting validity assessment
remains an open challenge.

Revisability gives this visibility an operational role. An analyst can respond
to a questionable assumption by changing the relevant definition or plan and
regenerating the affected artifacts, rather than accepting or rejecting the
entire report. Process evidence therefore supports informed judgment: it shows
what should be examined and where an intervention can be made. This perspective
also reframes trust as the ability to assess and respond to an output, rather
than confidence in the output alone.

\subsection{Human--AI Collaboration as Expertise Integration}

The staged workflow divides analytical work according to the strengths of the
system and the analyst. The system performs repetitive data inspection, code
generation, execution, artifact collection, and structured checks. The analyst
contributes soccer-specific definitions, selects meaningful comparisons, and
judges whether a result supports a credible tactical interpretation. This
division does not restrict domain knowledge to the initial request. The
interaction records show that participants introduced it during problem
clarification, plan revision, and post-result refinement, with planning and
refinement serving as its principal entry points.

The task-level patterns further suggest that the required balance changes with
analytical demand. L3 tasks had lower completion and required more execution
rounds than the other levels, consistent with the additional coordination
needed to connect several event patterns into a tactical account. Because task
content and fixed order also differed across levels, this pattern should not be
treated as a controlled effect of difficulty. It nevertheless indicates that
support for complex analysis should strengthen opportunities to revise
definitions, plans, and interpretations rather than relying on greater
automation alone.

Together, these findings position stage-aware collaboration as a means of
making analytical work both revisable and open to domain judgment. Its value
lies not only in reducing manual effort, but also in providing explicit points
at which analysts can shape, inspect, and verify an analysis as it develops.

\section{Limitations and Future Work}

The formative and task-based studies provide focused evidence from 5 and 16
participants, respectively. Their experience as university-team
analysts or soccer-focused researchers was well aligned with the analytical
workflow, but it captures only part of the range of roles and organizational
settings in which soccer analysis is conducted. In addition, the task ratings
describe the 33 completed tasks and include ratings from only the 15 participants
with at least one completed task; they omit ratings for the 15 operationally incomplete tasks and
therefore do not represent P013's task-level experience. Future evaluations
can extend this evidence through larger and more diverse samples, including
professional and academy analysts, coaches, and technical staff, while
examining successful and unsuccessful experiences together.

The implementation and evaluation focused on structured JSON event data with a
Wyscout 2017 configuration, providing a consistent setting in which to examine
the stage-aware workflow. Data providers nevertheless differ in their schemas,
terminology, coverage, and interfaces, and these resources can change between
versions. Event logs can support only concepts represented directly or
derivable from their recorded fields, while sample-based profiling may not
expose rare structures. Future work can evaluate transfer across providers,
schema versions, competitions, data types, and sports. It can also examine
whether incorporating official provider documentation or maintained data
dictionaries into dataset understanding and planning improves task feasibility
and reduces data-resolution failures.

LLM-generated plans and code provide flexibility in translating natural-language
questions into executable analyses, but their outputs remain sensitive to
model assumptions and can carry shared blind spots across coding and review
agents. Claim--evidence links make the resulting process easier to inspect, but
traceability alone does not establish computational or soccer-analytical
correctness. Future work can combine model-based review with deterministic
checks, independent models, or expert assessment. Studies with known analytical
errors can then measure whether the system identifies the error, whether users
recognize the warning and revise the appropriate stage, and whether the final
report corrects the original problem.

\section{Conclusion}

Soccer data analysis requires analysts to translate domain questions into
definitions and computations whose assumptions and supporting evidence can be
examined, yet prompt-to-report LLM tools can obscure these decisions and make
problems difficult to identify and revise. This paper presented AI Soccer
Analyst, which organizes LLM-assisted analysis into the inspectable stages of
Data Understanding, Problem Definition, Structured Planning, Execution,
Evidence-Grounded Reporting, and Interaction and Refinement while preserving opportunities for analysts to contribute domain
knowledge. In the evaluation, 33 of 48 analytical tasks met the operational
completion criteria. Exploratory tests supported favorable
participant perceptions of completed-task output quality, task achievement,
reliability, and verifiability after Holm correction. Interaction records showed that
participants contributed domain knowledge when clarifying questions, reviewing
plans, and refining results. Together, these findings illustrate
how participants used stage-aware human--AI collaboration and suggest that they
perceived it as supporting access to complex analytical work while retaining
opportunities for involvement in consequential decisions. The approach offers
a design direction for analytical systems that combine automation with
sustained human judgment and produce analyses that are easier to inspect,
revise, and verify.

\begin{acks}
This work was financially supported by JST FOREST Program (JPMJFR26153613) and JSPS
KAKENHI (26H02478). The study was approved by the General Affairs Committee
of the Graduate School of Informatics, Nagoya University (approval no.
I26-18(I26-09-01)).
Generative AI tools were used to assist with manuscript writing, editing, and figure preparation.
The authors reviewed all AI-assisted content and take full responsibility for the work.
\end{acks}

\bibliographystyle{ACM-Reference-Format}
\bibliography{software}

@article{wilcoxon1945individual,
  author  = {Frank Wilcoxon},
  title   = {Individual Comparisons by Ranking Methods},
  journal = {Biometrics Bulletin},
  year    = {1945},
  volume  = {1},
  number  = {6},
  pages   = {80--83},
  doi     = {10.2307/3001968}
}

@article{holm1979simple,
  author  = {Sture Holm},
  title   = {A Simple Sequentially Rejective Multiple Test Procedure},
  journal = {Scandinavian Journal of Statistics},
  volume  = {6},
  number  = {2},
  pages   = {65--70},
  year    = {1979},
  url     = {https://www.jstor.org/stable/4615733}
}

@inproceedings{brewer2025coach,
  author    = {Mollie Brewer and Kevin Childs and Celeste Wilkins and Spencer Thomas and Kristy Elizabeth Boyer and Jennifer A. Nichols and Kevin R. B. Butler and Garrett F. Beatty and Daniel P. Ferris},
  title     = {Coach, Data Analyst, and Protector: Exploring Data Practices of Collegiate Coaching Staff},
  booktitle = {Proceedings of the 2025 CHI Conference on Human Factors in Computing Systems},
  year      = {2025},
  publisher = {Association for Computing Machinery},
  address   = {New York, NY, USA},
  pages     = {1--13},
  doi       = {10.1145/3706598.3714026}
}

@inproceedings{lee2025crafting,
  author    = {Hanbyeol Lee and Erica Kleinman and Namsub Kim and Sangbeom Park and Casper Harteveld and Byungjoo Lee},
  title     = {Crafting Champions: An Observation Study of Esports Coaching Processes},
  booktitle = {Proceedings of the 2025 CHI Conference on Human Factors in Computing Systems},
  year      = {2025},
  publisher = {Association for Computing Machinery},
  address   = {New York, NY, USA},
  articleno = {991},
  numpages  = {20},
  doi       = {10.1145/3706598.3713141}
}

@inproceedings{weng2025bridging,
  author    = {Jian-Jia Weng and Calvin Ku and Jo Chien Wang and Chih-Jen Cheng and Tica Lin and Yu-An Su and Tsung-Hsun Tsai and You-Yi Lin and Lun-Wei Ku and Hung-Kuo Chu and Min-Chun Hu},
  title     = {Bridging Coaching Knowledge and {AI} Feedback to Enhance Motor Learning in Basketball Shooting Mechanics Through a Knowledge-Based {SOP} Framework},
  booktitle = {Proceedings of the 2025 CHI Conference on Human Factors in Computing Systems},
  year      = {2025},
  publisher = {Association for Computing Machinery},
  address   = {New York, NY, USA},
  doi       = {10.1145/3706598.3713324}
}

@inproceedings{goldi2025efficient,
  author    = {Andreas G{\"o}ldi and Roman Rietsche and Lyle H. Ungar},
  title     = {Efficient Management of {LLM}-Based Coaching Agents' Reasoning While Maintaining Interaction Quality and Speed},
  booktitle = {Proceedings of the 2025 CHI Conference on Human Factors in Computing Systems},
  year      = {2025},
  publisher = {Association for Computing Machinery},
  address   = {New York, NY, USA},
  articleno = {992},
  numpages  = {18},
  doi       = {10.1145/3706598.3713606}
}

@inproceedings{jorke2025gptcoach,
  author    = {Matthew J{\"o}rke and Shardul Sapkota and Lyndsea Warkenthien and Niklas Vainio and Paul Schmiedmayer and Emma Brunskill and James A. Landay},
  title     = {{GPTCoach}: Towards {LLM}-Based Physical Activity Coaching},
  booktitle = {Proceedings of the 2025 CHI Conference on Human Factors in Computing Systems},
  year      = {2025},
  publisher = {Association for Computing Machinery},
  address   = {New York, NY, USA},
  articleno = {993},
  numpages  = {46},
  doi       = {10.1145/3706598.3713819}
}

@inproceedings{song2025bleacherbot,
  author    = {Kyusik Kim and Hyungwoo Song and Jeongwoo Ryu and Changhoon Oh and Bongwon Suh},
  title     = {{BleacherBot}: {AI} Agent as a Sports Co-Viewing Partner},
  booktitle = {Proceedings of the 2025 CHI Conference on Human Factors in Computing Systems},
  year      = {2025},
  publisher = {Association for Computing Machinery},
  address   = {New York, NY, USA},
  articleno = {988},
  numpages  = {31},
  doi       = {10.1145/3706598.3714178}
}

@inproceedings{hirano2026solecoach,
  author    = {Toshihiro Hirano and Hitoshi Yoshihara and Yichen Peng and Chen-Chieh Liao and Erwin Wu and Hideki Koike},
  title     = {{SoleCoach}: Sole Pressure and {IMU}-Based {MLLMs} for Skill Coaching},
  booktitle = {Proceedings of the 2026 CHI Conference on Human Factors in Computing Systems},
  year      = {2026},
  publisher = {Association for Computing Machinery},
  address   = {New York, NY, USA},
  articleno = {1429},
  numpages  = {17},
  doi       = {10.1145/3772318.3791181}
}

@inproceedings{lee2026ballpark,
  author    = {Dokyung Lee and Jaeseong Ju and Hyungwoo Song and Hyunwoo Park},
  title     = {From Ballpark to Society: Understanding Stakeholders' Adaptation to Automated Judgment via {ABS} in Baseball},
  booktitle = {Proceedings of the 2026 CHI Conference on Human Factors in Computing Systems},
  year      = {2026},
  publisher = {Association for Computing Machinery},
  address   = {New York, NY, USA},
  articleno = {1426},
  doi       = {10.1145/3772318.3791249}
}

@inproceedings{brewer2026performance,
  author    = {Mollie Brewer and Kevin Childs and Spencer Thomas and Celeste Wilkins and Zachary R. Smith and Kristy Elizabeth Boyer and Jennifer A. Nichols and Kevin R. B. Butler and Garrett F. Beatty and Daniel P. Ferris},
  title     = {Improve my Performance, Protect my State of Mind: How Student-Athletes Engage with their Sports Data},
  booktitle = {Proceedings of the 2026 CHI Conference on Human Factors in Computing Systems},
  year      = {2026},
  publisher = {Association for Computing Machinery},
  address   = {New York, NY, USA},
  articleno = {1428},
  doi       = {10.1145/3772318.3791745}
}

@inproceedings{chen2026badminsense,
  author    = {Taizhou Chen and Kai Chen and Xingyu Liu and Pingchuan Ke and Zhida Sun},
  title     = {{BadminSense}: Enabling Fine-Grained Badminton Strokes Evaluation on Single Smartwatch},
  booktitle = {Proceedings of the 2026 CHI Conference on Human Factors in Computing Systems},
  year      = {2026},
  publisher = {Association for Computing Machinery},
  address   = {New York, NY, USA},
  articleno = {1424},
  numpages  = {20},
  doi       = {10.1145/3772318.3790998}
}

@inproceedings{amershi2019guidelines,
  author    = {Saleema Amershi and Dan Weld and Mihaela Vorvoreanu and Adam Fourney and Besmira Nushi and Penny Collisson and Jina Suh and Shamsi Iqbal and Paul N. Bennett and Kori Inkpen and Jaime Teevan and Ruth Kikin-Gil and Eric Horvitz},
  title     = {Guidelines for Human--{AI} Interaction},
  booktitle = {Proceedings of the 2019 CHI Conference on Human Factors in Computing Systems},
  year      = {2019},
  publisher = {Association for Computing Machinery},
  address   = {New York, NY, USA},
  articleno = {3},
  numpages  = {13},
  doi       = {10.1145/3290605.3300233}
}

@inproceedings{horvitz1999mixed,
  author    = {Eric Horvitz},
  title     = {Principles of Mixed-Initiative User Interfaces},
  booktitle = {Proceedings of the SIGCHI Conference on Human Factors in Computing Systems},
  year      = {1999},
  publisher = {Association for Computing Machinery},
  address   = {New York, NY, USA},
  pages     = {159--166},
  doi       = {10.1145/302979.303030}
}

@article{heer2019agency,
  author  = {Jeffrey Heer},
  title   = {Agency Plus Automation: Designing Artificial Intelligence into Interactive Systems},
  journal = {Proceedings of the National Academy of Sciences},
  volume  = {116},
  number  = {6},
  pages   = {1844--1850},
  year    = {2019},
  doi     = {10.1073/pnas.1807184115}
}

@article{shneiderman2020hcai,
  author  = {Ben Shneiderman},
  title   = {Human-Centered Artificial Intelligence: Reliable, Safe \& Trustworthy},
  journal = {International Journal of Human--Computer Interaction},
  volume  = {36},
  number  = {6},
  pages   = {495--504},
  year    = {2020},
  doi     = {10.1080/10447318.2020.1741118}
}

@article{parasuraman2000automation,
  author  = {Raja Parasuraman and Thomas B. Sheridan and Christopher D. Wickens},
  title   = {A Model for Types and Levels of Human Interaction with Automation},
  journal = {IEEE Transactions on Systems, Man, and Cybernetics---Part A: Systems and Humans},
  volume  = {30},
  number  = {3},
  pages   = {286--297},
  year    = {2000},
  doi     = {10.1109/3468.844354}
}

@article{lee2004trust,
  author  = {John D. Lee and Katrina A. See},
  title   = {Trust in Automation: Designing for Appropriate Reliance},
  journal = {Human Factors},
  volume  = {46},
  number  = {1},
  pages   = {50--80},
  year    = {2004},
  doi     = {10.1518/hfes.46.1.50.30392}
}

@inproceedings{bansal2021whole,
  author    = {Gagan Bansal and Tongshuang Wu and Joyce Zhou and Raymond Fok and Besmira Nushi and Ece Kamar and Marco Tulio Ribeiro and Daniel S. Weld},
  title     = {Does the Whole Exceed Its Parts? The Effect of {AI} Explanations on Complementary Team Performance},
  booktitle = {Proceedings of the 2021 CHI Conference on Human Factors in Computing Systems},
  year      = {2021},
  publisher = {Association for Computing Machinery},
  address   = {New York, NY, USA},
  articleno = {81},
  numpages  = {16},
  doi       = {10.1145/3411764.3445717}
}

@article{pappalardo2019wyscout,
  author  = {Luca Pappalardo and Paolo Cintia and Paolo Ferragina and Emanuele Massucco and Dino Pedreschi and Fosca Giannotti},
  title   = {A Public Data Set of Spatio-Temporal Match Events in Soccer Competitions},
  journal = {Scientific Data},
  volume  = {6},
  number  = {1},
  pages   = {236},
  year    = {2019},
  doi     = {10.1038/s41597-019-0247-7}
}

@software{aider,
  author  = {Paul Gauthier},
  title   = {Aider: AI Pair Programming in Your Terminal},
  year    = {2024},
  url     = {https://github.com/Aider-AI/aider}
}

@article{kandel2012enterprise,
  author  = {Sean Kandel and Andreas Paepcke and Joseph M. Hellerstein and Jeffrey Heer},
  title   = {Enterprise Data Analysis and Visualization: An Interview Study},
  journal = {IEEE Transactions on Visualization and Computer Graphics},
  volume  = {18},
  number  = {12},
  pages   = {2917--2926},
  year    = {2012},
  doi     = {10.1109/TVCG.2012.219}
}

@article{ragan2016provenance,
  author  = {Eric D. Ragan and Alex Endert and Jibonananda Sanyal and Jian Chen},
  title   = {Characterizing Provenance in Visualization and Data Analysis: An Organizational Framework of Provenance Types and Purposes},
  journal = {IEEE Transactions on Visualization and Computer Graphics},
  volume  = {22},
  number  = {1},
  pages   = {31--40},
  year    = {2016},
  doi     = {10.1109/TVCG.2015.2467551}
}

@article{bucinca2021trust,
  author  = {Zana Bu{\c c}inca and Maja Barbara Malaya and Krzysztof Z. Gajos},
  title   = {To Trust or to Think: Cognitive Forcing Functions Can Reduce Overreliance on {AI} in {AI}-Assisted Decision-Making},
  journal = {Proceedings of the ACM on Human-Computer Interaction},
  volume  = {5},
  number  = {CSCW1},
  year    = {2021},
  articleno = {188},
  numpages = {21},
  doi     = {10.1145/3449287}
}

@incollection{ehsan2020hcxai,
  author    = {Upol Ehsan and Mark O. Riedl},
  title     = {Human-Centered Explainable {AI}: Towards a Reflective Sociotechnical Approach},
  booktitle = {HCI International 2020---Late Breaking Papers: Multimodality and Intelligence},
  publisher = {Springer International Publishing},
  address   = {Cham},
  pages     = {449--466},
  year      = {2020},
  doi       = {10.1007/978-3-030-60117-1_33}
}

@article{braun2006thematic,
  author  = {Virginia Braun and Victoria Clarke},
  title   = {Using Thematic Analysis in Psychology},
  journal = {Qualitative Research in Psychology},
  volume  = {3},
  number  = {2},
  pages   = {77--101},
  year    = {2006},
  doi     = {10.1191/1478088706qp063oa}
}

@article{yeung2023interpretable,
  author  = {Calvin C. K. Yeung and Rory Bunker and Keisuke Fujii},
  title   = {A Framework of Interpretable Match Results Prediction in Football with {FIFA} Ratings and Team Formation},
  journal = {PLOS ONE},
  volume  = {18},
  number  = {4},
  pages   = {e0284318},
  year    = {2023},
  doi     = {10.1371/journal.pone.0284318}
}

@article{yeung2024onevsone,
  author  = {Calvin Yeung and Keisuke Fujii},
  title   = {A Strategic Framework for Optimal Decisions in Football 1-vs-1 Shot-Taking Situations: An Integrated Approach of Machine Learning, Theory-Based Modeling, and Game Theory},
  journal = {Complex \& Intelligent Systems},
  volume  = {10},
  pages   = {5989--6008},
  year    = {2024},
  doi     = {10.1007/s40747-024-01466-4}
}

@article{yeung2024tactics,
  author  = {Calvin Yeung and Rory Bunker and Keisuke Fujii},
  title   = {Unveiling Multi-Agent Strategies: A Data-Driven Approach for Extracting and Evaluating Team Tactics from Football Event and Freeze-Frame Data},
  journal = {Journal of Robotics and Mechatronics},
  volume  = {36},
  number  = {3},
  pages   = {603--617},
  year    = {2024},
  doi     = {10.20965/jrm.2024.p0603}
}

@article{yeung2025nmstpp,
  author  = {Calvin Yeung and Tony Sit and Keisuke Fujii},
  title   = {Transformer-Based Neural Marked Spatio Temporal Point Process Model for Analyzing Football Match Events},
  journal = {Applied Intelligence},
  volume  = {55},
  articleno = {335},
  year    = {2025},
  doi     = {10.1007/s10489-024-05996-9}
}

@article{yeung2025openstarlab,
  author        = {Calvin Yeung and Kenjiro Ide and Taiga Someya and Keisuke Fujii},
  title         = {{OpenSTARLab}: open approach for spatio-temporal agent data analysis in soccer},
  journal   = {Complex \& Intelligent Systems},
  volume    = {11},
  articleno = {342},
  year      = {2025},
  doi       = {10.1007/s40747-025-01965-y}
}

@article{openai2025gptoss,
  author        = {{OpenAI}},
  title         = {{gpt-oss-120b \& gpt-oss-20b} Model Card},
  year          = {2025},
  eprint        = {2508.10925},
  archivePrefix = {arXiv},
  primaryClass  = {cs.CL},
  doi           = {10.48550/arXiv.2508.10925}
}

@inproceedings{kwon2023efficient,
  author    = {Kwon, Woosuk and Li, Zhuohan and Zhuang, Siyuan and Sheng, Ying and Zheng, Lianmin and Yu, Cody Hao and Gonzalez, Joseph E. and Zhang, Hao and Stoica, Ion},
  title     = {Efficient Memory Management for Large Language Model Serving with {PagedAttention}},
  booktitle = {Proceedings of the 29th Symposium on Operating Systems Principles},
  year      = {2023},
  pages     = {611--626},
  doi       = {10.1145/3600006.3613165}
}

@misc{vercel2026nextjs,
  author       = {{Vercel}},
  title        = {{Next.js} Documentation},
  year         = {2026},
  url          = {https://nextjs.org/docs},
  urldate      = {2026-07-15}
}

@misc{fastapi2026docs,
  author       = {{FastAPI}},
  title        = {{FastAPI} Documentation},
  year         = {2026},
  url          = {https://fastapi.tiangolo.com/},
  urldate      = {2026-07-15}
}

@misc{docker2026overview,
  author       = {{Docker, Inc.}},
  title        = {What Is Docker?},
  year         = {2026},
  url          = {https://docs.docker.com/get-started/docker-overview/},
  urldate      = {2026-07-15}
}

@inproceedings{drosos2024rubberduck,
  author    = {Ian Drosos and Advait Sarkar and Xiaotong Xu and Carina Negreanu and Sean Rintel and Lev Tankelevitch},
  title     = {{"It's Like a Rubber Duck That Talks Back"}: Understanding Generative {AI}-Assisted Data Analysis Workflows through a Participatory Prompting Study},
  booktitle = {Proceedings of the 3rd Annual Meeting of the Symposium on Human-Computer Interaction for Work},
  year      = {2024},
  publisher = {Association for Computing Machinery},
  address   = {New York, NY, USA},
  articleno = {16},
  numpages  = {21},
  doi       = {10.1145/3663384.3663389}
}

@inproceedings{gu2024planning,
  author    = {Ken Gu and Madeleine Grunde-McLaughlin and Andrew McNutt and Jeffrey Heer and Tim Althoff},
  title     = {How Do Data Analysts Respond to {AI} Assistance? A Wizard-of-Oz Study},
  booktitle = {Proceedings of the 2024 CHI Conference on Human Factors in Computing Systems},
  year      = {2024},
  publisher = {Association for Computing Machinery},
  address   = {New York, NY, USA},
  numpages  = {22},
  doi       = {10.1145/3613904.3641891}
}

@inproceedings{gu2024verification,
  author    = {Ken Gu and Ruoxi Shang and Tim Althoff and Chenglong Wang and Steven Mark Drucker},
  title     = {How Do Analysts Understand and Verify {AI}-Assisted Data Analyses?},
  booktitle = {Proceedings of the 2024 CHI Conference on Human Factors in Computing Systems},
  year      = {2024},
  publisher = {Association for Computing Machinery},
  address   = {New York, NY, USA},
  articleno = {748},
  numpages  = {22},
  doi       = {10.1145/3613904.3642497}
}

@inproceedings{shih2024cellsync,
  author    = {Jasmine Y. Shih and Vishal Mohanty and Yannis Katsis and Hariharan Subramonyam},
  title     = {Leveraging Large Language Models to Enhance Domain Expert Inclusion in Data Science Workflows},
  booktitle = {Extended Abstracts of the 2024 CHI Conference on Human Factors in Computing Systems},
  year      = {2024},
  publisher = {Association for Computing Machinery},
  address   = {New York, NY, USA},
  numpages  = {11},
  doi       = {10.1145/3613905.3651115}
}

@inproceedings{guo2024agency,
  author    = {Jiajing Guo and Vikram Mohanty and Jorge Henrique Piazentin Ono and Hongtao Hao and Liang Gou and Liu Ren},
  title     = {Investigating Interaction Modes and User Agency in Human--{LLM} Collaboration for Domain-Specific Data Analysis},
  booktitle = {Extended Abstracts of the 2024 CHI Conference on Human Factors in Computing Systems},
  year      = {2024},
  publisher = {Association for Computing Machinery},
  address   = {New York, NY, USA},
  articleno = {203},
  numpages  = {9},
  doi       = {10.1145/3613905.3651042}
}

@inproceedings{kim2026laps,
  author    = {Jaehoon Kim and Dayoung Jeong and Beejin Son and Hansung Kim and Bogoan Kim and Kyungsik Han},
  title     = {{LAPS}: Automating Hypothesis-Driven Statistical Analysis of Public Survey Using Large Language Models},
  booktitle = {Proceedings of the 2026 CHI Conference on Human Factors in Computing Systems},
  year      = {2026},
  publisher = {Association for Computing Machinery},
  address   = {New York, NY, USA},
  numpages  = {19},
  doi       = {10.1145/3772318.3791665}
}

@inproceedings{wu2022aichains,
  author    = {Tongshuang Wu and Michael Terry and Carrie J. Cai},
  title     = {{AI} Chains: Transparent and Controllable Human--{AI} Interaction by Chaining Large Language Model Prompts},
  booktitle = {Proceedings of the 2022 CHI Conference on Human Factors in Computing Systems},
  year      = {2022},
  publisher = {Association for Computing Machinery},
  address   = {New York, NY, USA},
  numpages  = {22},
  doi       = {10.1145/3491102.3517582}
}

@inproceedings{wu2022promptchainer,
  author    = {Tongshuang Wu and Ellen Jiang and Aaron Donsbach and Jeff Gray and Alejandra Molina and Michael Terry and Carrie J. Cai},
  title     = {{PromptChainer}: Chaining Large Language Model Prompts through Visual Programming},
  booktitle = {Extended Abstracts of the 2022 CHI Conference on Human Factors in Computing Systems},
  year      = {2022},
  publisher = {Association for Computing Machinery},
  address   = {New York, NY, USA},
  numpages  = {10},
  doi       = {10.1145/3491101.3519729}
}

@article{wang2024dataformulator,
  author  = {Chenglong Wang and John Thompson and Bongshin Lee},
  title   = {Data Formulator: {AI}-Powered Concept-Driven Visualization Authoring},
  journal = {IEEE Transactions on Visualization and Computer Graphics},
  volume  = {30},
  number  = {1},
  pages   = {1128--1138},
  year    = {2024},
  doi     = {10.1109/TVCG.2023.3326585}
}

@inproceedings{xie2024waitgpt,
  author    = {Liwenhan Xie and Chengbo Zheng and Haijun Xia and Huamin Qu and Zhu-Tian Chen},
  title     = {{WaitGPT}: Monitoring and Steering Conversational {LLM} Agent in Data Analysis with On-the-Fly Code Visualization},
  booktitle = {Proceedings of the 37th Annual ACM Symposium on User Interface Software and Technology},
  year      = {2024},
  publisher = {Association for Computing Machinery},
  address   = {New York, NY, USA},
  numpages  = {14},
  doi       = {10.1145/3654777.3676374}
}

@inproceedings{vaithilingam2024dynavis,
  author    = {Priyan Vaithilingam and Elena L. Glassman and Jeevana Priya Inala and Chenglong Wang},
  title     = {{DynaVis}: Dynamically Synthesized {UI} Widgets for Visualization Editing},
  booktitle = {Proceedings of the 2024 CHI Conference on Human Factors in Computing Systems},
  year      = {2024},
  publisher = {Association for Computing Machinery},
  address   = {New York, NY, USA},
  articleno = {985},
  numpages  = {17},
  doi       = {10.1145/3613904.3642639}
}

@inproceedings{kazemitabaar2024steering,
  author    = {Majeed Kazemitabaar and Jack Williams and Ian Drosos and Tovi Grossman and Austin Z. Henley and Carina Negreanu and Advait Sarkar},
  title     = {Improving Steering and Verification in {AI}-Assisted Data Analysis with Interactive Task Decomposition},
  booktitle = {Proceedings of the 37th Annual ACM Symposium on User Interface Software and Technology},
  year      = {2024},
  publisher = {Association for Computing Machinery},
  address   = {New York, NY, USA},
  numpages  = {19},
  doi       = {10.1145/3654777.3676345}
}

@inproceedings{zhou2026confirmation,
  author    = {Jieyu Zhou and Aryan Roy and Sneh Gupta and Daniel Weitekamp and Christopher J. MacLellan},
  title     = {When Should Users Check? Modeling Confirmation Frequency in Multi-Step Agentic {AI} Tasks},
  booktitle = {Proceedings of the 2026 CHI Conference on Human Factors in Computing Systems},
  year      = {2026},
  publisher = {Association for Computing Machinery},
  address   = {New York, NY, USA},
  articleno = {1649},
  numpages  = {20},
  doi       = {10.1145/3772318.3790655}
}

@inproceedings{tankelevitch2024metacognitive,
  author    = {Lev Tankelevitch and Viktor Kewenig and Auste Simkute and Ava Elizabeth Scott and Advait Sarkar and Abigail Sellen and Sean Rintel},
  title     = {The Metacognitive Demands and Opportunities of Generative {AI}},
  booktitle = {Proceedings of the 2024 CHI Conference on Human Factors in Computing Systems},
  year      = {2024},
  publisher = {Association for Computing Machinery},
  address   = {New York, NY, USA},
  doi       = {10.1145/3613904.3642902}
}

\clearpage
\onecolumn
\appendix
\section{Administered Surveys and Follow-Up Interview Protocol}
\label{app:survey}

This appendix presents the study instruments in three parts. Survey~1 (S1) examines
participants' soccer-analysis backgrounds, current workflows, and expectations
for AI-supported analysis. The formative follow-up interview clarifies and
extends participants' Survey~1 responses, with particular attention to
verification practices and appropriate boundaries for automation. Survey~2
(S2) evaluates the system after tasks at three levels of analytical complexity: L1
(descriptive retrieval), L2 (comparative interpretation), and L3 (tactical
synthesis). It concludes with an overall assessment of the user experience.
Stable identifiers
\textbf{S1-Q1}--\textbf{S1-Q14} and
\textbf{S2-Q1}--\textbf{S2-Q13} correspond to the numbered survey items. An
asterisk marks a required response.

\subsection{Survey 1: Formative Evaluation}

\surveysection{Purpose and administration}{Survey 1}

The participant-facing introduction explains that the survey is part of a
formative study of an AI-supported soccer-analysis system. Its purpose is to
collect expert input for future design and development, including desired
analytical functions and forms of support. Responses are used only for
research, are not published in personally identifiable form, and may inform
system design and subsequent research.

\surveysection{Participant identification and background}{Required items}

\surveyitem{S1-Q1*}{Email address}
\responseformat{Free text. Participants are instructed to use the address
listed on their consent form; it is collected for research communication and
to match the response to the consent record.}

\surveyitem{S1-Q2*}{How many years of experience do you have in soccer
analysis?}
\responseformat{Select one: less than 1 year; 1--3 years; 4--6 years;
7--10 years; 10 years or more.}

\surveysection{Current analysis workflow}{Required items}

\surveyitem{S1-Q3*}{How do you currently conduct soccer analysis?}
\responseformat{Select all that apply.}
\begin{itemize}[leftmargin=2em,nosep]
  \item Analysis conducted mainly through manual work.
  \item Manual analysis with some automated steps.
  \item Analysis combining manual work and AI assistance.
  \item Analysis conducted mainly through automation or AI assistance.
\end{itemize}

\surveyitem{S1-Q4*}{Which aspects of your current analysis workflow are the
most difficult?}
\responseformat{Select all that apply: data preparation; discovering useful
insights; validation and validity checking; report preparation; sharing and
communicating results; time required; other.}

\surveyitem{S1-Q5*}{Current-workflow challenges}
\responseformat{Rate each statement.}
\ratingscale{strongly disagree}{strongly agree}
\begin{center}
\small
\begin{tabularx}{0.94\linewidth}{@{}Xccccc@{}}
  \toprule
  Statement & 1 & 2 & 3 & 4 & 5 \\
  \midrule
  Verification of analytical results requires substantial manual work.
    & $\circ$ & $\circ$ & $\circ$ & $\circ$ & $\circ$ \\
  Existing analysis tools provide insufficient support for inspecting or
  verifying the process and evidence leading to a result.
    & $\circ$ & $\circ$ & $\circ$ & $\circ$ & $\circ$ \\
  The current analysis workflow is time-consuming.
    & $\circ$ & $\circ$ & $\circ$ & $\circ$ & $\circ$ \\
  Repeated manual operations reduce workflow efficiency.
    & $\circ$ & $\circ$ & $\circ$ & $\circ$ & $\circ$ \\
  \bottomrule
\end{tabularx}
\end{center}

\surveysection{Expectations for AI-supported analysis}{Required item}

\surveyitem{S1-Q6*}{Expectations for the system}
\responseformat{Rate each statement.}
\ratingscale{strongly disagree}{strongly agree}
\begin{center}
\small
\begin{tabularx}{0.94\linewidth}{@{}Xccccc@{}}
  \toprule
  Statement & 1 & 2 & 3 & 4 & 5 \\
  \midrule
  Making intermediate reasoning processes visible would increase trust in
  AI-generated analyses.
    & $\circ$ & $\circ$ & $\circ$ & $\circ$ & $\circ$ \\
  Transparency and verifiability are important in AI-supported analysis.
    & $\circ$ & $\circ$ & $\circ$ & $\circ$ & $\circ$ \\
  An LLM-based system is expected to reduce the burden of manual work.
    & $\circ$ & $\circ$ & $\circ$ & $\circ$ & $\circ$ \\
  An AI-supported workflow is expected to increase the speed of analysis.
    & $\circ$ & $\circ$ & $\circ$ & $\circ$ & $\circ$ \\
  \bottomrule
\end{tabularx}
\end{center}

\surveysection{Semi-structured written responses}{Applicable items}

The form explains that these items are open-ended and may be followed by one
or two brief questions when clarification is needed. Follow-ups are intended
to clarify the response while minimizing participant burden.

\surveyitem{S1-Q7*}{Please describe your current analysis workflow.}
\responseformat{Open text.}

\surveyitem{S1-Q8*}{What are the main challenges in your current workflow?}
\responseformat{Open text.}

\surveyitem{S1-Q9*}{Which analysis tasks require the most manual work?}
\responseformat{Open text.}

\surveyitem{S1-Q10*}{At which stage of the analysis process are corrections or
rework most likely to occur, and why?}
\responseformat{Open text.}

\surveyitem{S1-Q11*}{How do you currently check or validate the validity of
analytical results?}
\responseformat{Open text.}

\surveyitem{S1-Q12*}{What do you expect from an LLM-based analysis-support
system?}
\responseformat{Open text.}

\surveyitem{S1-Q13*}{What elements would be necessary to increase your trust
in an AI-supported analysis system?}
\responseformat{Open text.}

\surveyitem{S1-Q14*}{What concerns would you have when using such a system?}
\responseformat{Open text.}

\subsection{Formative Follow-Up Interview}

The follow-up interview is a separate, semi-structured protocol rather than a
numbered component of either survey form. Interview questions are selected in
response to the participant's Survey~1 answers. They are not administered as a fixed
checklist: each category has a specific diagnostic purpose, and the
interviewer asks only the interview questions needed to understand the
participant's account.

\begin{description}[leftmargin=0pt,style=nextline,itemsep=0.8em]
  \item[Rating rationale.]
  \emph{Purpose:} explain unusually high or low ratings and distinguish the
  participant's interpretation of a scale item from its intended construct.
  \emph{Interview questions:} Ask why the participant selected the response and request a
  concrete example from a recent analysis task.

  \item[Workflow reconstruction.]
  \emph{Purpose:} connect general descriptions to the steps and tools used in
  actual practice. \emph{Interview question:} Ask how the participant usually performs
  that step.

  \item[Verification and rework.]
  \emph{Purpose:} identify the evidence analysts inspect and the failures that
  initiate revision. \emph{Interview questions:} Ask what information the participant
  inspects when checking a result and what commonly causes an analysis to be
  revised.

  \item[Automation boundaries.]
  \emph{Purpose:} locate decisions that can be automated and decisions where
  domain input has high analytical leverage. \emph{Interview question:} Ask when an
  automated action would be acceptable without confirmation and when the
  system should stop and ask.

  \item[Domain conflict and responsibility.]
  \emph{Purpose:} understand how analysts resolve disagreement between
  generated evidence and soccer knowledge. \emph{Interview question:} Ask how the
  participant would decide what to do if a system result conflicted with their
  soccer knowledge.

  \item[Coverage check.]
  \emph{Purpose:} surface practices or concerns not anticipated by the survey.
  \emph{Interview question:} Ask whether the questionnaire omitted anything important
  about workflow, trust, verification, or the use of AI.
\end{description}

Short probes such as asking why, requesting an example, or asking how a step
is usually performed are used to clarify vague answers, apparent
contradictions, or distinctive practices.

\subsection{Survey 2: User Evaluation}

\surveysection{Purpose and administration}{Survey 2}

The participant-facing introduction explains that Survey~2 evaluates the
AI-supported soccer-analysis system after use. Participants assess both their
experience and the generated analytical results. The survey contains three
task-level sections, completed after L1, L2, and L3 respectively, followed by
one overall-system evaluation section. Responses are used only for research,
are not published in personally identifiable form, and may inform system
evaluation and future development.

\surveysection{Participant identification and background}{Required items}

\surveyitem{S2-Q1*}{Email address}
\responseformat{Free text. Participants are instructed to use the address
listed on their consent form; it is collected for research communication and
to match the response to the consent record.}

\surveyitem{S2-Q2*}{How many years of experience do you have in soccer
analysis?}
\responseformat{Select one: less than 1 year; 1--3 years; 4--6 years;
7--10 years; 10 years or more.}

\surveysection{Part 1: L1 descriptive-retrieval task evaluation}{Required items}

\surveyitem{S2-Q3*}{Enter the name of the L1 job executed in the system.}
\responseformat{Free text.}

\surveyitem{S2-Q4*}{Rate the system for the L1 task.}
\responseformat{One rating per criterion.}
\taskratingmatrix

\surveysection{Part 2: L2 comparative-interpretation task evaluation}{Required items}

\surveyitem{S2-Q5*}{Enter the name of the L2 job executed in the system.}
\responseformat{Free text.}

\surveyitem{S2-Q6*}{Rate the system for the L2 task.}
\responseformat{One rating per criterion.}
\taskratingmatrix

\surveysection{Part 3: L3 tactical-synthesis task evaluation}{Required items}

\surveyitem{S2-Q7*}{Enter the name of the L3 job executed in the system.}
\responseformat{Free text.}

\surveyitem{S2-Q8*}{Rate the system for the L3 task.}
\responseformat{One rating per criterion.}
\taskratingmatrix

\surveysection{Part 4: Overall system evaluation}{Required item}

\surveyitem{S2-Q9*}{Overall system evaluation}
\responseformat{Rate each statement.}
\ratingscale{strongly disagree}{strongly agree}
\begin{center}
\small
\begin{tabularx}{0.96\linewidth}{@{}Xccccc@{}}
  \toprule
  Statement & 1 & 2 & 3 & 4 & 5 \\
  \midrule
  I could trust the analytical results generated by the system.
    & $\circ$ & $\circ$ & $\circ$ & $\circ$ & $\circ$ \\
  I could judge whether the system output was valid or problematic.
    & $\circ$ & $\circ$ & $\circ$ & $\circ$ & $\circ$ \\
  The system made it easy to inspect and verify the process leading to an
  analytical result.
    & $\circ$ & $\circ$ & $\circ$ & $\circ$ & $\circ$ \\
  The system made it easy to inspect and verify the evidence supporting an
  analytical result.
    & $\circ$ & $\circ$ & $\circ$ & $\circ$ & $\circ$ \\
  The system helped me progress through analysis tasks efficiently.
    & $\circ$ & $\circ$ & $\circ$ & $\circ$ & $\circ$ \\
  The system reduced unnecessary manual work.
    & $\circ$ & $\circ$ & $\circ$ & $\circ$ & $\circ$ \\
  Dialogue with the system, including questions, corrections, and feedback,
  helped me progress with the analysis.
    & $\circ$ & $\circ$ & $\circ$ & $\circ$ & $\circ$ \\
  I would like to use this system in future soccer-analysis work.
    & $\circ$ & $\circ$ & $\circ$ & $\circ$ & $\circ$ \\
  \bottomrule
\end{tabularx}
\end{center}

\surveysection{Overall written feedback}{Optional items}

\surveyitem{S2-Q10}{What aspects of the system were good?}
\responseformat{Open text.}

\surveyitem{S2-Q11}{While using the system, what was difficult, hard to
understand, or inefficient?}
\responseformat{Open text.}

\surveyitem{S2-Q12}{Did any output feel untrustworthy? If so, please explain.}
\responseformat{Open text.}

\surveyitem{S2-Q13}{What information or explanation would you need to verify
the results more effectively?}
\responseformat{Open text.}

\clearpage
\section{Supplementary Formative Study Results}
\label{app:formative-analysis}

\subsection{Study Data and Participants}

This appendix provides supporting descriptive results and coding details for
the formative study reported in Section~\ref{sec:formative-study}. The
semi-structured protocol combined the pre-interview survey reproduced in
Appendix~\ref{app:survey} with follow-up questions. Administrative fields and
personally identifying information were excluded.

{Participants were recruited through email invitations. Eligibility
required at least 1 year of soccer-analysis experience in a professional
context or multiple years of experience in a collegiate context. In total, 7
eligible individuals were contacted, 5 agreed to participate and completed
the survey, and 2 of those 5 also completed a follow-up interview.
Follow-up interviews were conducted selectively rather than with every
participant. The 2 participants were contacted because their open-ended
responses raised points that required further clarification or elaboration;
the remaining survey responses were sufficiently detailed for the formative
analysis.
Participants received compensation of approximately US\$20. The survey took
approximately 30 minutes to complete. The follow-up interviews were conducted
by email, took approximately 15 minutes, and focused primarily on the
open-ended survey questions. Follow-up responses were incorporated as
extensions to the corresponding open-ended survey answers, producing the
qualitative dataset used for coding.}

{The study included 5 survey participants, and all percentages
use $n=5$.}
Qualitative responses could receive multiple codes; consequently, theme
percentages do not sum to 100\%. Participants are identified using the anonymized labels
E1--E5; these labels distinguish participants only and do not indicate rank or
level of expertise.

\subsection{Participant and Workflow Context}

\paragraph{Soccer-analysis experience.}
Table~\ref{tab:formative-experience} summarizes the participants' experience.
All participants had at least 1 year of experience, and 4 had 4 or
more years of experience. The responses therefore reflect sustained practical
exposure to soccer analysis rather than first impressions from novice users.

\begin{table}[htbp]
  
  \centering
  \small
  \caption{Participant roles, work contexts, and soccer-analysis experience.}
  \label{tab:formative-experience}
  \begin{tabularx}{0.92\linewidth}{@{}lXl@{}}
    \toprule
    Participant & Role and work context & Experience \\
    \midrule
    E1 & Professional soccer analyst & 4--6 years \\
    E2 & Professional soccer analyst & 1--3 years \\
    E3 & Professional soccer analyst & 4--6 years \\
    E4 & Collegiate soccer analyst & 7--10 years \\
    E5 & Professional soccer analyst & 7--10 years \\
    \bottomrule
  \end{tabularx}
\end{table}

\paragraph{Current workflow.}

S1-Q3 allowed multiple selections; accordingly, the percentages in
Table~\ref{tab:formative-workflow} do not sum to 100\%. The overlapping
selections indicate that participants combined workflow modes rather than
working exclusively manually or automatically. Participants who selected
mainly automated or AI-assisted analysis still described substantial human
work in interpretation, checking, and communication, suggesting that existing
automation shifted effort rather than removing it.

\begin{table}[htbp]
  \centering
  \small
  \caption{Current soccer-analysis workflow modes.}
  \label{tab:formative-workflow}
  \begin{tabularx}{0.86\linewidth}{@{}Xrr@{}}
    \toprule
    Workflow mode & Count & Percentage \\
    \midrule
    Mainly automation or AI assistance & 4 & 80\% \\
    Manual analysis with some automated steps & 2 & 40\% \\
    Manual work combined with AI assistance & 2 & 40\% \\
    Mainly manual work & 0 & 0\% \\
    \bottomrule
  \end{tabularx}
\end{table}

\paragraph{Workflow difficulties.}

The workflow-difficulty item (S1-Q4) permitted multiple selections; the
reported difficulties are shown in Table~\ref{tab:formative-difficulties}. No
predefined difficulty dominated the sample; instead, the selections span data
preparation, insight discovery, validation, communication, and time cost. This
distribution suggests that workflow burden is distributed across analytical
stages rather than confined to a single bottleneck. Report preparation received no selections, whereas E1 identified manual
report creation and repeated adjustment of graph and slide layouts as
difficult. This contrast illustrates the additional context contributed by the
follow-up questions.

\begin{table}[htbp]
  \centering
  \small
  \caption{Difficulties in the current analytical workflow.}
  \label{tab:formative-difficulties}
  \begin{tabularx}{0.82\linewidth}{@{}Xrr@{}}
    \toprule
    Difficulty & Count & Percentage \\
    \midrule
    Data preparation & 2 & 40\% \\
    Discovering useful insights & 2 & 40\% \\
    Validation and validity checking & 2 & 40\% \\
    Sharing and communicating results & 2 & 40\% \\
    Time required & 2 & 40\% \\
    Real-time data acquisition (other) & 1 & 20\% \\
    Report preparation & 0 & 0\% \\
    \bottomrule
  \end{tabularx}
\end{table}

\paragraph{Current-workflow ratings.}

Ratings range from 1 (strongly disagree) to 5 (strongly agree). The
interquartile range (IQR) is the 25th--75th percentile interval.
Table~\ref{tab:formative-current-ratings} reports the response distributions
and descriptive statistics. Time cost received the highest median rating,
while ratings of manual verification burden and process opacity varied across
participants. These results identify efficiency as the clearest shared concern
while indicating that verification needs depend on participants' existing
workflows.

\begin{table}[htbp]
  \centering
  \small
  \setlength{\tabcolsep}{4pt}
  \caption{Ratings of the current analytical workflow.}
  \label{tab:formative-current-ratings}
  \begin{tabularx}{\linewidth}{@{}Xccc@{}}
    \toprule
    S1-Q5 item & Distribution (1/2/3/4/5) & Median & IQR \\
    \midrule
    Verification requires substantial manual work & 1/0/2/2/0 & 3 & 3--4 \\
    Existing tools insufficiently expose process and evidence & 0/2/1/0/2 & 3 & 2--5 \\
    Current workflow is time-consuming & 1/1/0/3/0 & 4 & 2--4 \\
    Repeated manual operations reduce efficiency & 1/1/2/0/1 & 3 & 2--3 \\
    \bottomrule
  \end{tabularx}
\end{table}

\paragraph{Expectations for AI-supported analysis.}
Table~\ref{tab:formative-expectation-ratings} summarizes participants' initial
expectations concerning trust, verifiability, workload, and speed. Every rating
was 4 or 5, with the strongest consensus concerning transparency and
verifiability, reduced manual work, and greater speed. The ratings therefore
indicate agreement with the initial assumptions guiding the formative study:
AI assistance should improve efficiency while supporting verifiability and
appropriately calibrated trust.

\begin{table}[htbp]
  \centering
  \small
  \setlength{\tabcolsep}{4pt}
  \caption{Expectations for AI-supported analysis.}
  \label{tab:formative-expectation-ratings}
  \begin{tabularx}{\linewidth}{@{}Xccc@{}}
    \toprule
    S1-Q6 item & Distribution (1/2/3/4/5) & Median & IQR \\
    \midrule
    Visible intermediate reasoning would increase trust & 0/0/0/2/3 & 5 & 4--5 \\
    Transparency and verifiability are important & 0/0/0/1/4 & 5 & 5--5 \\
    An LLM system could reduce manual work & 0/0/0/1/4 & 5 & 5--5 \\
    AI support could increase analysis speed & 0/0/0/1/4 & 5 & 5--5 \\
    \bottomrule
  \end{tabularx}
\end{table}

\subsection{Qualitative Coding and Codebook}

Written responses to S1-Q7--S1-Q14, including their incorporated follow-up
clarifications, were segmented into meaningful units. A response could receive
more than one code. The first pass applied three deductive categories:

\begin{itemize}[leftmargin=*]
  \item \textbf{Trust and calibrated reliance:} conditions affecting whether
  an AI-supported result can be relied upon appropriately.
  \item \textbf{Verifiability and evidence:} practices or needs for inspecting
  assumptions, data, process, and outputs.
  \item \textbf{Efficiency and workload:} time, repeated manual operations,
  bottlenecks, and desired workload reductions.
\end{itemize}

A second pass compared uncaptured units across participants and formed
inductive codes. Similar codes were merged, their boundaries were clarified,
and prevalence was counted once per participant per code. This was an
exploratory single-coder analysis; no intercoder-agreement claim is made.
Table~\ref{tab:formative-codebook} presents the final codebook and participant
prevalence. Its operational definitions state the observable content used to
apply each code consistently.

\begin{table}[htbp]
  \centering
  \small
  \setlength{\tabcolsep}{3pt}
  \caption{Final deductive--inductive codebook and participant prevalence.}
  \label{tab:formative-codebook}
  \begin{tabularx}{\linewidth}{@{}p{0.10\linewidth}p{0.32\linewidth}Xrr@{}}
    \toprule
    Type & Code & Operational definition & Participants & Prevalence \\
    \midrule
    Deductive & Trust and calibrated reliance & Interpretability, literacy, contextual fit, or uncertainty affecting appropriate reliance & 4 & 80\% \\
    Deductive & Verifiability and evidence & Inspection of assumptions, primary data, distributions, video, alternative sources, or process records & 4 & 80\% \\
    Deductive & Efficiency and workload & Time-consuming acquisition, aggregation, reporting, checking, or repeated manual work & 4 & 80\% \\
    Inductive & Data access, timeliness, and infrastructure & Real-time availability, APIs, provider constraints, cloud readiness, storage, and pipeline access & 4 & 80\% \\
    Inductive & Human interpretation and domain alignment & Need to connect outputs with soccer knowledge, tactical judgment, video, and local team criteria & 4 & 80\% \\
    Inductive & Communication and actionable outputs & Reporting, prioritization, staff-facing presentation, self-service access, and information overload & 4 & 80\% \\
    Inductive & Organizational adoption and capability distribution & Literacy, staff engagement, individual dependency, continuity, and uneven technical capability & 3 & 60\% \\
    Inductive & Automation boundaries and skill preservation & Work suitable for automation versus work requiring judgment, plus risks of deskilling & 2 & 40\% \\
    Inductive & Data security and information governance & Risk of sensitive information leakage or insecure use of AI-supported tools & 2 & 40\% \\
    \bottomrule
  \end{tabularx}
\end{table}

\clearpage
\clearpage
\section{Supplementary User Evaluation Results}
\label{app:evaluation-analysis}

\subsection{Study Data and Participants}

This appendix reports the detailed cohort definition, descriptive results,
inferential tests, system-behavior summaries, and qualitative
codebook for the task-based evaluation described in
Section~\ref{sec:evaluation-method}.

Participants were recruited through email invitations sent to university
soccer-team analysts, master's or doctoral students conducting soccer-focused
research, and individuals belonging to both groups. In total, 23 individuals
were recruited, of whom 16 completed the experimental procedure and submitted
Survey~2. Participants received compensation of approximately US\$20. Each task
was designed to take approximately 45--60 minutes.

The retained study data comprised the 16 Survey~2 responses and system records
for 48 linked analytical tasks, including task questions, job states, system
logs, intermediate artifacts, and generated reports where available.
All counts use only the retained analytical tasks and surveys. 

\subsection{Cohort and Analysis Boundaries}

The survey workbook contained 16 responses. All 16 participants and
their 48 linked analytical tasks were included. P013's three tasks were
operationally incomplete, but the participant remained in the cohort so that
operationally incomplete tasks and overall-system perceptions were represented. The analysis contains
exactly 1 analytical task per participant at each level. Table~\ref{tab:evaluation-experience}
summarizes the participants' reported experience.

\begin{table}[htbp]
  \centering
  \small
  \caption{Soccer-analysis experience in the evaluation cohort ($n=16$).}
  \label{tab:evaluation-experience}
  \begin{tabular}{@{}lrr@{}}
    \toprule
    Experience & Participants & Percentage \\
    \midrule
    Less than 1 year & 3 & 18.8\% \\
    1--3 years & 9 & 56.3\% \\
    4--6 years & 2 & 12.5\% \\
    10 years or more & 2 & 12.5\% \\
    \bottomrule
  \end{tabular}
\end{table}

Task-rating analyses include only the 33 tasks whose final top-level status was
\texttt{complete}: 12 at L1, 11 at L2, and 10 at L3. Fifteen participants
had at least one completed task and contributed to the participant-aggregated
analysis; P013 had none and contributed no completed-task rating. Overall-system
ratings use a single response from each of all 16 participants.

\subsection{Question Types, Operationally Incomplete Tasks, and System Behavior}

Each analytical request received one primary analytical-demand category.
Tactical-synthesis tasks were the most common, while comparison and
ranking/filtering tasks had the lowest descriptive completion rates
(Table~\ref{tab:evaluation-question-types}). These categories describe the
requests and are not inferred participant intentions.

\begin{table}[htbp]
  \centering
  \small
  \caption{Primary analytical demands among the 48 analytical tasks.}
  \label{tab:evaluation-question-types}
  \begin{tabular}{@{}lrrrr@{}}
    \toprule
    Question type & Tasks & Participants & Complete & Completion \\
    \midrule
    Tactical synthesis & 22 & 15 & 17 & 77.3\% \\
    Descriptive & 10 & 10 & 8 & 80.0\% \\
    Comparison & 4 & 4 & 2 & 50.0\% \\
    Ranking/filtering & 5 & 5 & 2 & 40.0\% \\
    Temporal & 3 & 3 & 2 & 66.7\% \\
    Relationship/modeling & 4 & 3 & 2 & 50.0\% \\
    \bottomrule
  \end{tabular}
\end{table}

Causes of incompletion were derived from retained job states, execution reviews, and
artifact status. In total, 4 tasks across 2 participants were infeasible with the
available event data because they required tracking or off-ball information;
all three of P013's tasks were operationally incomplete for this reason. Six tasks across 5 participants failed because
execution could not resolve required dataset fields, identifiers, tags, or
available records. Another 3 tasks stopped because the system ran out of memory
during the analysis, 1 failed because the required report was
saved to the wrong path, and 1 was cancelled after reaching the experiment's time limit (24
hours).

Table~\ref{tab:evaluation-system-behavior} summarizes execution rounds,
duration, and token use across the three task levels.

\begin{table}[htbp]
  \centering
  \small
  \setlength{\tabcolsep}{5pt}
  \caption{System-behavior descriptives by task level. Each measure includes
  all 16 analytical tasks at that level; IQR is reported as Q1--Q3.}
  \label{tab:evaluation-system-behavior}
  \begin{tabular}{@{}lrrlrrrr@{}}
    \toprule
    Level & Tasks & Complete & Measure & Mean & SD & Median & IQR (Q1--Q3) \\
    \midrule
    L1 & 16 & 12 & Execution rounds & 9.75 & 14.54 & 3.00 & 2.00--11.25 \\
       &    &    & Duration (s) & 2,700.6 & 4,315.2 & 978.0 & 571.8--3,195.5 \\
       &    &    & Tokens & 26,895 & 12,174 & 27,674 & 14,850--38,252 \\
    \addlinespace
    L2 & 16 & 11 & Execution rounds & 12.25 & 17.55 & 3.50 & 1.75--16.25 \\
       &    &    & Duration (s) & 1,934.6 & 1,911.6 & 1,033.5 & 740.0--2,387.0 \\
       &    &    & Tokens & 80,974 & 145,809 & 47,959 & 23,985--74,975 \\
    \addlinespace
    L3 & 16 & 10 & Execution rounds & 11.00 & 8.17 & 9.50 & 7.75--13.00 \\
       &    &    & Duration (s) & 3,413.9 & 2,778.8 & 2,255.0 & 1,828.3--3,325.5 \\
       &    &    & Tokens & 54,777 & 52,578 & 43,336 & 29,921--62,688 \\
    \bottomrule
  \end{tabular}
\end{table}

The medians showed that a typical L3 task required more execution rounds than a
typical L1 or L2 task. The means were higher than the medians for several
measures, especially L2 token use, indicating that a small number of
resource-intensive tasks increased the averages. These descriptive differences
do not establish that task level caused greater resource use because no
between-level test was conducted.
Durations are wall-clock measures that can include time awaiting confirmation
and delays due to resource contention; they are not active task time. Token
counts likewise describe system activity rather than analytical quality.
\subsection{Completed-Task Ratings}

Table~\ref{tab:evaluation-task-ratings} combines the descriptive and
inferential results for the 4 task-rating items, each measured on a 5-point
scale. For the primary analysis, completed-task scores were averaged within
participant so that each participant contributed 1 value to each overall
outcome. This accounts for ratings from the same participant not being
independent and prevents participants with more completed tasks from receiving
greater weight. The level-specific rows use the single completed-task rating
from each participant at that level. Consequently, $n=15$ for the overall rows,
whereas $n=12$, 11, and 10 for L1, L2, and L3, respectively. All results are
conditioned on operational completion.

Ratings were summarized descriptively using medians and
interquartile ranges. Exploratory exact one-sided Wilcoxon signed-rank tests
were used to determine whether the ratings tended to exceed the neutral scale
value of 3 without assuming a normal distribution. The test treats participants as independent and assumes
that the nonzero differences are approximately symmetric; under this
assumption, the tested location corresponds to the median difference. Exact
inference was used because the samples were small. For each outcome, the
analysis calculated the difference between each included value and 3. The
hypotheses were:
\begin{itemize}
  \item $H_0$: The location of these differences is less than or equal to 0.
  \item $H_1$: The location of these differences is greater than 0.
\end{itemize}
Differences equal to 0 were omitted,
and tied absolute differences received their average rank. The positive-rank
sum, $W^+$, was the sum of ranks associated with values above 3, and the 
one-sided $p$-value was the probability, under $H_0$, of obtaining a positive-rank
sum at least as large as the observed value. We report both the
unadjusted exact $p$-values and Holm-adjusted $p$-values. Testing several related
outcomes increases the probability of obtaining at least one false-positive
result. Holm correction~\cite{holm1979simple} was therefore used to control the
family-wise error rate while retaining the individual outcome tests. It was
applied separately across the four primary participant-aggregated outcomes and
across the twelve secondary level-by-outcome tests because these families answer
different questions: overall perceptions of completed outputs and patterns
within task-complexity levels, respectively. At $\alpha=.05$, the unadjusted and
Holm significance columns show the conclusions before and after correction.
The tests within each level evaluate ratings against 3 and do not compare the
three levels with one another.

\begin{table*}[htbp]
  \centering
  \small
  \setlength{\tabcolsep}{3pt}
  \caption{Completed-task rating descriptives and exploratory
  exact one-sided Wilcoxon signed-rank tests versus the neutral value of 3.
  Adjusted $p$-values use Holm correction across the four participant-aggregated
  outcomes and, separately, across the twelve level-by-outcome tests.}
  \label{tab:evaluation-task-ratings}
  \begin{tabular*}{\textwidth}{@{\extracolsep{\fill}}llrrrrrrrcc@{}}
    \toprule
    Level & Outcome & $n$ & Mean & SD & Median & $W^+$ & Raw $p$ &
    \shortstack{Adjusted\\$p$} &
    \shortstack{Unadjusted\\significant} &
    \shortstack{Holm\\significant} \\
    \midrule
    Overall & Output quality & 15 & 4.111 & 0.821 & 4.000 & 78.0 & .0002 & .0007 & Yes & Yes \\
       & Task achievement & 15 & 4.344 & 0.638 & 4.500 & 105.0 & .0001 & .0002 & Yes & Yes \\
       & Reliability & 15 & 3.922 & 0.936 & 4.000 & 96.0 & .0022 & .0044 & Yes & Yes \\
       & Verifiability & 15 & 4.056 & 1.011 & 4.333 & 95.5 & .0022 & .0044 & Yes & Yes \\
    \addlinespace
    L1 & Output quality & 12 & 4.083 & 1.084 & 4.500 & 43.0 & .0078 & .0469 & Yes & Yes \\
       & Task achievement & 12 & 4.333 & 0.985 & 5.000 & 63.5 & .0024 & .0195 & Yes & Yes \\
       & Reliability & 12 & 3.833 & 1.193 & 4.000 & 38.0 & .0312 & .0850 & Yes & No \\
       & Verifiability & 12 & 4.083 & 1.505 & 5.000 & 62.0 & .0283 & .0850 & Yes & No \\
    \addlinespace
    L2 & Output quality & 11 & 4.182 & 0.751 & 4.000 & 45.0 & .0020 & .0176 & Yes & Yes \\
       & Task achievement & 11 & 4.364 & 0.505 & 4.000 & 66.0 & .0005 & .0059 & Yes & Yes \\
       & Reliability & 11 & 4.273 & 0.647 & 4.000 & 55.0 & .0010 & .0098 & Yes & Yes \\
       & Verifiability & 11 & 4.364 & 0.505 & 4.000 & 66.0 & .0005 & .0059 & Yes & Yes \\
    \addlinespace
    L3 & Output quality & 10 & 4.100 & 1.287 & 5.000 & 41.0 & .0137 & .0684 & Yes & No \\
       & Task achievement & 10 & 4.400 & 0.966 & 5.000 & 52.5 & .0049 & .0342 & Yes & Yes \\
       & Reliability & 10 & 3.800 & 1.476 & 4.500 & 29.0 & .0859 & .0859 & No & No \\
       & Verifiability & 10 & 4.000 & 1.054 & 4.000 & 33.5 & .0195 & .0781 & Yes & No \\
    \bottomrule
  \end{tabular*}
\end{table*}

For the participant-aggregated analysis, all four outcomes met
the unadjusted threshold and remained significant after Holm correction, with
adjusted $p$-values from .0002 to .0044. Among the twelve secondary
level-by-outcome tests, eleven met the unadjusted threshold, but seven remained
significant after correction: output quality and task achievement at L1; all
four outcomes at L2; and task achievement at L3. Thus, the correction changed
the inferential classification of L1 reliability and verifiability and L3
output quality and verifiability. These tests do not establish differences
among levels. All interpretations apply only to operationally completed tasks
and do not include the experience of operationally incomplete tasks.

\subsection{Overall System Ratings}

Overall-system results include all 16 participants, including P013, and use a
single response from each participant. For each item, the exact one-sided Wilcoxon signed-rank test used
$H_0$ and $H_1$ as follows:
\begin{itemize}
  \item $H_0$: The response location is less than or equal to 3.
  \item $H_1$: The response location is greater than 3.
\end{itemize}
Differences from 3 equal to 0 were
omitted, tied absolute differences received average ranks, and $W^+$ summed
the ranks for responses above 3. The tests were treated as
exploratory. As with the completed-task outcomes, testing multiple related items
increases the chance of at least one false-positive result. We therefore report
both unadjusted exact one-sided $p$-values and Holm-adjusted $p$-values, with
Holm correction applied across the eight overall-system items as a separate
family. The unadjusted and Holm significance columns show the conclusions before
and after correction at $\alpha=.05$.

\begin{table*}[htbp]
  \centering
  \small
  \setlength{\tabcolsep}{3pt}
  \caption{Overall-system rating descriptives and exploratory
  exact one-sided signed-rank tests versus 3 ($n=16$ for each item). Adjusted
  $p$-values use Holm correction across the eight items.}
  \label{tab:evaluation-overall-tests}
  \begin{tabular*}{\textwidth}{@{\extracolsep{\fill}}llrrrrrrrcc@{}}
    \toprule
    Level & Outcome & $n$ & Mean & SD & Median & $W^+$ & Raw $p$ &
    \shortstack{Adjusted\\$p$} &
    \shortstack{Unadjusted\\significant} &
    \shortstack{Holm\\significant} \\
    \midrule
    Overall & Overall trust & 16 & 3.688 & 1.250 & 4.000 & 72.5 & .0333 & .1333 & Yes & No \\
       & Judge validity & 16 & 3.562 & 1.413 & 4.000 & 75.0 & .0873 & .1333 & No & No \\
       & Process verifiable & 16 & 3.875 & 1.148 & 4.000 & 58.0 & .0107 & .0645 & Yes & No \\
       & Evidence verifiable & 16 & 3.562 & 1.031 & 3.500 & 37.5 & .0391 & .1333 & Yes & No \\
       & Efficient progress & 16 & 3.875 & 1.310 & 4.000 & 87.5 & .0146 & .0729 & Yes & No \\
       & Reduced manual work & 16 & 4.438 & 1.094 & 5.000 & 110.5 & .0006 & .0049 & Yes & Yes \\
       & Helpful interaction & 16 & 4.000 & 1.033 & 4.000 & 93.5 & .0026 & .0184 & Yes & Yes \\
       & Future use & 16 & 3.625 & 1.204 & 3.500 & 45.0 & .0449 & .1333 & Yes & No \\
    \bottomrule
  \end{tabular*}
\end{table*}

Seven overall-system items met the unadjusted $p<.05$ threshold,
but only reduced manual work (adjusted $p=.0049$) and helpful interaction
(adjusted $p=.0184$) remained significant after Holm correction. The other five
items that met the unadjusted threshold did not retain that classification after
family-wise error control. Descriptively,
reduced manual work had the highest mean (4.438), followed by helpful interaction
(4.000). Judge validity and evidence verifiability had the lowest means (both
3.562). Evidence verifiability met the unadjusted threshold ($p=.0391$) but not
the Holm-adjusted threshold (adjusted $p=.1333$), while judge validity did not
meet the unadjusted threshold ($p=.0873$).

\subsection{Qualitative Analysis and Theme Definitions}

Free-text responses were organized by their source question: positive aspects,
difficulties, untrustworthy outputs, and additional verification needs.
Responses were segmented into meaningful units and assigned one or more
descriptive labels, which were consolidated into four broad themes. Two themes
capture positive experiences, and two capture improvement needs. Counts in
Table~\ref{tab:evaluation-codebook} are unique participants mentioning each
theme; a participant could contribute to more than one theme. Responses stating
no specific trust issue or verification need were retained in the audit data
but were not counted as evidence for an improvement theme.  The analysis was
conducted by a single analyst; no intercoder-agreement claim is made.

\begin{table}[htbp]
  \centering
  \small
  \setlength{\tabcolsep}{3pt}
  \caption{Open-response themes and participant prevalence.}
  \label{tab:evaluation-codebook}
  \begin{tabularx}{\linewidth}{@{}p{0.15\linewidth}p{0.27\linewidth}Xr@{}}
    \toprule
    Orientation & Theme & Definition & $n$ \\
    \midrule
    Positive & Analytical efficiency and accessibility & Faster access to
    analysis, automation, reporting, and use without programming & 9 \\
    Positive & Analytical support and new insights & Planning, suggestions,
    evidence, and perspectives that support analytical decisions & 6 \\
    Improvement & Interaction continuity and recovery & Support for continuing,
    revising, or recovering an analysis when interaction or execution breaks
    down & 11 \\
    Improvement & Analytical credibility and verification & Support for judging
    analytical credibility through transparent evidence and corroboration & 11 \\
    \bottomrule
  \end{tabularx}
\end{table}

The four-theme structure treats narrow, low-frequency observations as evidence
within broader themes rather than as standalone findings. The two positive
themes explain the favorable assessments of efficiency, accessibility, and
analytical support. The two improvement themes motivate persistent refinement
and active verification; considered with the analysis of operationally incomplete tasks, they also
support the need for reusable data-provider knowledge. Counts are descriptive
of this retained cohort and should not be interpreted as population
prevalence.

\clearpage
\section{Agent, Stage, and Substage Prompt Templates}
\label{app:agent-prompts}

This appendix documents the prompts associated with the four
agent roles and the five LLM-mediated stages of AI Soccer Analyst's six-stage
workflow. Execution comprises two substages---Analysis Execution and
Execution Review---with a separate prompt for each; consequently, six stage
and substage prompt templates are presented. The templates are shortened from
the implemented prompts to emphasize each role's responsibility, required
context, constraints, and output contract. Angle-bracketed terms denote runtime
values supplied for a particular job. The Data Understanding stage is excluded
because it uses a deterministic profiling function rather than an LLM prompt.

\lstdefinestyle{agentprompt}{
  backgroundcolor=\color{surveylight},
  frame=single,
  rulecolor=\color{surveyaccent},
  basicstyle=\ttfamily\scriptsize,
  breaklines=true,
  breakatwhitespace=false,
  columns=fullflexible,
  keepspaces=true,
  showstringspaces=false,
  xleftmargin=0.5em,
  xrightmargin=0.5em,
  aboveskip=0.8em,
  belowskip=0.8em
}

\subsection{Agent System Prompts}

\subsubsection{AnalystAgent}\leavevmode\par
\begin{lstlisting}[style=agentprompt]
You are a Soccer Analyst specializing in clear, evidence-based soccer
analysis. Analyze soccer problems through match context, tactics, player
performance, team behavior, and game-state dynamics. Identify relevant
soccer concepts, organize observations, and explain analytical implications
in a structured way. Do not invent data, events, or conclusions beyond the
user's request. Keep every output concise, factual, structured, and verifiable.
\end{lstlisting}

\subsubsection{CoderAgent}\leavevmode\par
\begin{lstlisting}[style=agentprompt]
You are the CoderAgent for an isolated soccer analytics worker. Implement the
confirmed analytical plan as transparent, executable Python. Treat supplied
dataset paths as read-only runtime inputs, edit only the managed analysis
script unless helper files are required, and write all report-facing outputs
to the managed results directory. Do not invent schema details or wait for
interactive input. Record results, validations, diagnostics, and output paths
in execution_result.json.
\end{lstlisting}

\subsubsection{ReviewAgent}\leavevmode\par
ReviewAgent uses the same system prompt as AnalystAgent. Its review-specific
behavior is defined by the Execution Review substage prompt
presented below.

\subsubsection{InteractionRefinementAgent}\leavevmode\par
\begin{lstlisting}[style=agentprompt]
You are a Soccer Analyst handling post-report interaction and refinement.
Answer questions from existing report and result artifacts when possible.
When feedback requires a change, route it to the earliest pipeline stage that
must be updated and describe the affected artifacts. Ask the user only when
the intended entity, metric, scope, or preference remains ambiguous. Do not
invent unsupported data or claims. Return strict JSON only.
\end{lstlisting}

\subsection{Stage and Substage Prompt Templates}

\subsubsection{Problem Definition---AnalystAgent}\leavevmode\par
\begin{lstlisting}[style=agentprompt]
Define the user's soccer analytics problem as a normalized JSON object.

User request:
<USER_REQUEST>

Use only information stated by the user. Preserve the requested objective,
entities, scope, constraints, and expected result. Do not invent datasets,
metrics, events, hardware, or limitations. Ask only questions that must be
resolved before planning; otherwise return an empty question list.

Return only valid JSON:
{
  "task": "<specific analytical objective or null>",
  "limitations": ["<user-provided limitation>"],
  "clarifying_questions": ["<required question>"]
}
\end{lstlisting}

\subsubsection{Structured Planning---AnalystAgent}\leavevmode\par
\begin{lstlisting}[style=agentprompt]
Create an executable and user-readable soccer analytics plan.

Problem definition:
<PROBLEM_DEFINITION_JSON>

Dataset understanding:
<DATASET_SUMMARY>

Additional retrieved dataset context:
<RETRIEVED_DATASET_CONTEXT>

Use only supported files, fields, values, identifiers, and mappings. Include an
early schema/value preflight before filtering or aggregation. Make every step
concrete, include validations for empty or inconsistent results, and request a
targeted dataset-detail command when essential context is missing.

Return only valid JSON with:
{
  "title": "<plan title>",
  "goal_for_user": "<plain-language goal>",
  "assumptions": ["<assumption>"],
  "inputs_needed": ["<full dataset path>"],
  "human_readable_steps": [
    {"step": 1, "name": "<name>", "explanation": "<explanation>"}
  ],
  "tool_plan": [
    {
      "step": 1,
      "objective": "<objective>",
      "actions": ["<action>"],
      "data_requirements": ["<requirement>"],
      "outputs": ["<artifact or value>"],
      "validation": ["<check>"]
    }
  ],
  "expected_outputs": ["<output>"],
  "open_questions": ["<question>"]
}
\end{lstlisting}

\subsubsection{Execution: Analysis Execution---CoderAgent}\leavevmode\par
The template below is the task input that CoderAgent passes to Aider for an
execution round. It is not Aider's internal system prompt; Aider manages the
additional instructions used to operate its coding workflow.

\begin{lstlisting}[style=agentprompt]
Implement the confirmed soccer analytics plan in <ANALYSIS_SCRIPT>.

Plan:
<PLAN_JSON>

Dataset root:
<DATASET_PATH>

Managed results directory:
<OUTPUT_DIRECTORY>

Latest review and prior execution context:
<REVIEW_AND_EXECUTION_CONTEXT>

Read only inputs supported by the plan. Inspect actual record shapes and values
before filtering. Generate validated, non-empty analytical outputs in the
managed results directory. Never call input() or access secrets or unrelated
files. Write execution_result.json with status, results, diagnostics,
intermediate checks, validation metadata, and paths to every report-facing
artifact. During revision, fix the latest review issue while preserving
already-correct behavior.
\end{lstlisting}

\subsubsection{Execution: Execution Review---ReviewAgent}\leavevmode\par
\begin{lstlisting}[style=agentprompt]
Review the coding agent's soccer analytics execution.

Confirmed plan:
<PLAN_JSON>

Execution records and saved-artifact metadata:
<EXECUTION_AND_ARTIFACT_CONTEXT>

Decide whether the execution answers the analytical goal, follows the plan,
and has adequate evidence. Check failures, empty filters, result consistency,
CSV contents, and plot metadata. Never accept an execution merely because code
ran or files exist. Use user clarification only for intent or preference; use
dataset clarification for inspectable dataset facts; otherwise request a
focused coder revision.

Return only valid JSON:
{
  "status": "complete | needs_coder_revision |
             needs_user_clarification | needs_data_clarification",
  "feedback": "<actionable feedback or completion summary>",
  "clarifying_questions": ["<question>"],
  "data_json_path": "<one JSON path or null>",
  "result_files": ["<result path>"]
}
\end{lstlisting}

\subsubsection{Evidence-Grounded Reporting---AnalystAgent}\leavevmode\par
\begin{lstlisting}[style=agentprompt]
Write the final soccer analytics report in Markdown.

User question:
<USER_QUESTION>

Confirmed plan and reviewed results:
<PLAN_AND_RESULT_CONTEXT>

User-facing tables and image artifacts:
<REPORT_ARTIFACTS>

Cleaned Python processing sections:
<CODE_SECTIONS>

Answer the question directly from reviewed, user-facing outputs. Include the
most relevant values and embed every supplied image by file name. Attach at
most one evidence marker such as [S1] to a supported claim, and link that
marker to exactly one cleaned-code section. Do not expose internal pipeline
files or claim to have visually interpreted image pixels. Do not invent
numbers, entities, methods, or conclusions. Output only Markdown and end with
a concise Conclusion section.
\end{lstlisting}

\subsubsection{Interaction and Refinement---InteractionRefinementAgent}\leavevmode\par
\begin{lstlisting}[style=agentprompt]
Classify and respond to the user's post-report message.

Recent chat:
<CHAT_HISTORY>

User message:
<USER_FEEDBACK_OR_QUESTION>

Current task, plan, execution summary, report, and artifact metadata:
<CURRENT_ARTIFACT_CONTEXT>

Answer directly when existing artifacts support the response. Ask the user
only when the entity, metric, scope, or preference remains ambiguous. Propose
refinement only when an artifact must change, and route it to the earliest
responsible stage: problem-definition, plan, execution, or report.

Return only valid JSON:
{
  "action": "answer_only | ask_user | refine",
  "target_stage": "null | problem-definition | plan | execution | report",
  "response": "<user-facing response>",
  "artifact_updates": ["<intended update>"],
  "artifact_lookup_query": "<lookup query or null>",
  "rationale": "<routing reason>"
}
\end{lstlisting}

\clearpage
\section{Planning Retrieval Functions}
\label{app:planning-retrieval}

During Structured Planning, the AnalystAgent can request additional dataset
information before producing the analytical plan. The agent returns one
function call, the backend executes it, and the result is added to the planning
context. The agent then either requests further information or produces the
final plan.

\begin{table}[h]
\caption{Dataset retrieval functions available during Structured Planning.}
\label{tab:planning-retrieval-functions}
\begin{tabularx}{\textwidth}{@{}p{0.28\textwidth}X@{}}
\toprule
Function & Purpose \\
\midrule
\nolinkurl{get_group_summary} & Retrieves the structure and source files associated with a logical dataset group, such as events, matches, or players. \\
\nolinkurl{get_file_summary} & Retrieves the structure, field information, and observed values of a specific dataset file. \\
\nolinkurl{get_team_metadata} & Retrieves team metadata used to connect team names or abbreviations to dataset identifiers. \\
\nolinkurl{get_competition_metadata} & Retrieves competition metadata used to connect leagues, tournaments, or seasons to dataset identifiers. \\
\bottomrule
\end{tabularx}
\end{table}

These functions allow the AnalystAgent to ground its proposed coding steps in
the available data. The backend limits the number of retrieval rounds to
prevent an indefinite planning loop. If the required information remains
unavailable, planning ends with an explicit error rather than continuing with
unsupported assumptions.

\definecolor{uiaccent}{HTML}{11877F}
\definecolor{uibackground}{HTML}{F4F7F8}
\definecolor{uiborder}{HTML}{D6E0E5}
\definecolor{uimint}{HTML}{DFF5E6}
\definecolor{uigreen}{HTML}{087A3B}
\definecolor{uiuser}{HTML}{EAF6F5}
\definecolor{uiagent}{HTML}{F5F8F8}
\definecolor{uievidence}{HTML}{FFF4C4}
\definecolor{uiwhite}{HTML}{FFFFFF}
\definecolor{uicodecomment}{HTML}{4F7A62}
\definecolor{uicodestring}{HTML}{9A5B24}

\newcommand{\uicard}[2][uiwhite]{%
  \par\smallskip\noindent
  \fcolorbox{uiborder}{#1}{%
    \begin{minipage}{\dimexpr\linewidth-2\fboxsep-2\fboxrule\relax}
      #2
    \end{minipage}}%
  \par\smallskip}
\newcommand{\uistatus}[1]{%
  \colorbox{uimint}{\textcolor{uigreen}{\strut\footnotesize\bfseries #1}}}
\newcommand{\uievidenceid}[1]{%
  \colorbox{uibackground}{\textcolor{uiaccent}{\strut\footnotesize\bfseries #1}}}
\newcommand{\evidencecard}[4]{%
  \uicard[uiuser]{%
    \textcolor{uiaccent}{\bfseries Evidence #1}\par\smallskip
    #2}
  \uicard[uievidence]{%
    \textbf{Analysis section #3}\par\smallskip
    #4}}
\newcommand{\mdtwo}[1]{%
  \par\bigskip\noindent{\large\bfseries #1}\par\smallskip}
\newcommand{\mdthree}[1]{%
  \par\medskip\noindent{\bfseries #1}\par\smallskip}

\lstdefinestyle{uipython}{
  language=Python,
  backgroundcolor=\color{uiagent},
  frame=single,
  rulecolor=\color{uiborder},
  framesep=7pt,
  basicstyle=\ttfamily\tiny,
  keywordstyle=\color{uiaccent}\bfseries,
  commentstyle=\color{uicodecomment},
  stringstyle=\color{uicodestring},
  numbers=left,
  numberstyle=\scriptsize\color{gray},
  numbersep=8pt,
  showstringspaces=false,
  breaklines=true,
  breakatwhitespace=false,
  columns=fullflexible,
  keepspaces=true,
  tabsize=4,
  captionpos=b,
  literate={→}{{$\rightarrow$}}1 {‑}{{-}}1 {×}{{$\times$}}1 { }{{ }}1
}

\clearpage
\section{User Interface Overview}
\label{app:ui}

\begin{figure}[!ht]
  \centering
  \includegraphics[width=0.76\linewidth]{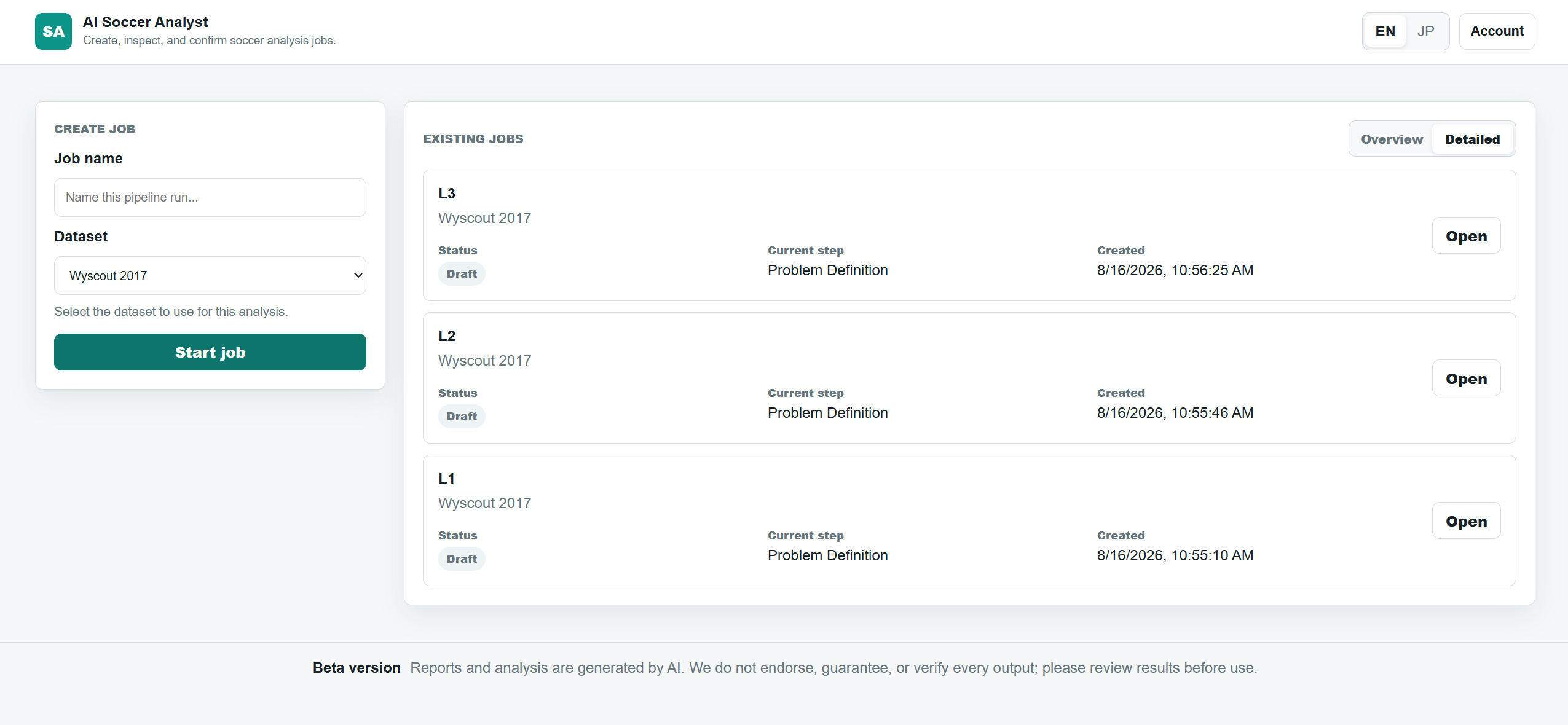}
  \caption{Job dashboard for creating an analysis and reopening existing jobs.}
  \label{fig:ui-home}
\end{figure}

\begin{figure}[!ht]
  \centering
  \includegraphics[width=0.76\linewidth]{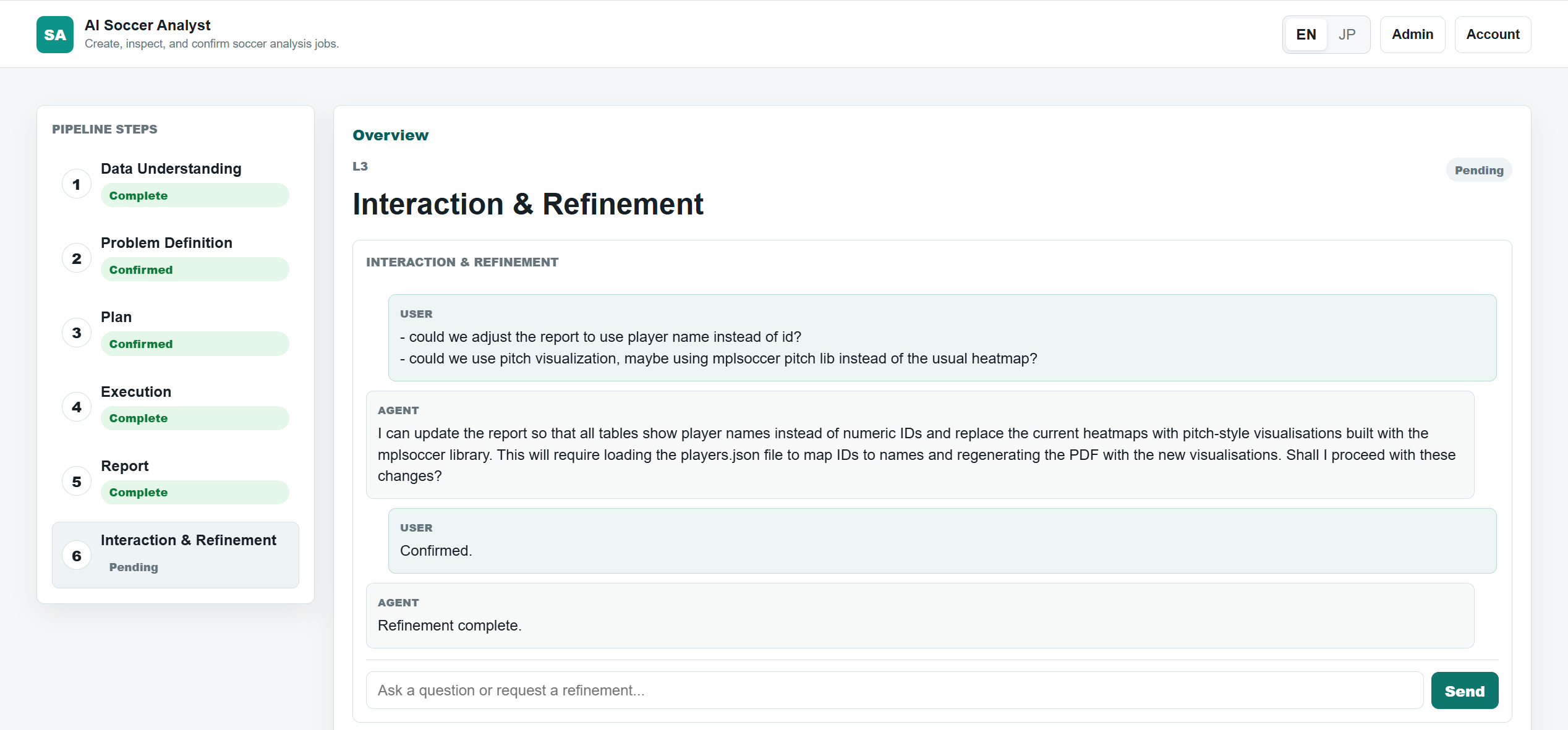}
  \caption{{Interaction and Refinement view illustrating
  refinement routing and user confirmation.}}
  \label{fig:ui-refine}
\end{figure}

The interface organizes the analysis around a job dashboard and a staged
workflow. The dashboard supports task creation and access to existing jobs,
whereas the workspace exposes the current stage, its user-facing artifact,
and the controls needed to confirm or revise it. The workspace
navigator represents the six analytical stages. Execution Review is an
internal substage of Execution rather than a separate stage, so it does not
appear as a separate navigation item. Figures~\ref{fig:ui-home}--
{\ref{fig:ui-refine} show the five principal views. The
refinement view illustrates the routing and confirmation flow.}

\begin{figure}[!ht]
  \centering
  \includegraphics[width=0.76\linewidth]{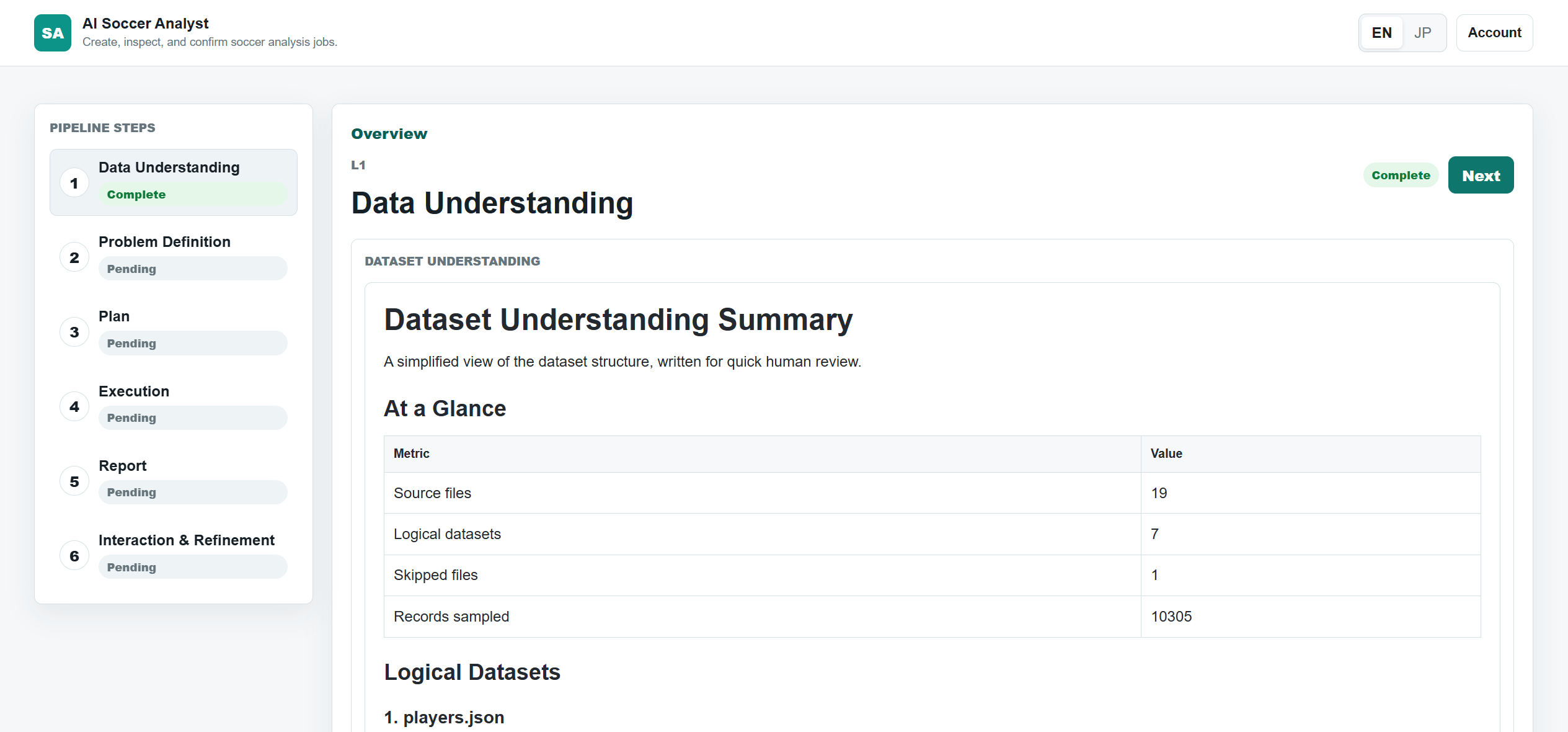}
  \caption{Analysis workspace showing the six-stage pipeline and the artifact
  for the currently selected stage.}
  \label{fig:ui-pipeline}
\end{figure}

\clearpage

\begin{figure}[!ht]
  \centering
  \includegraphics[width=0.76\linewidth]{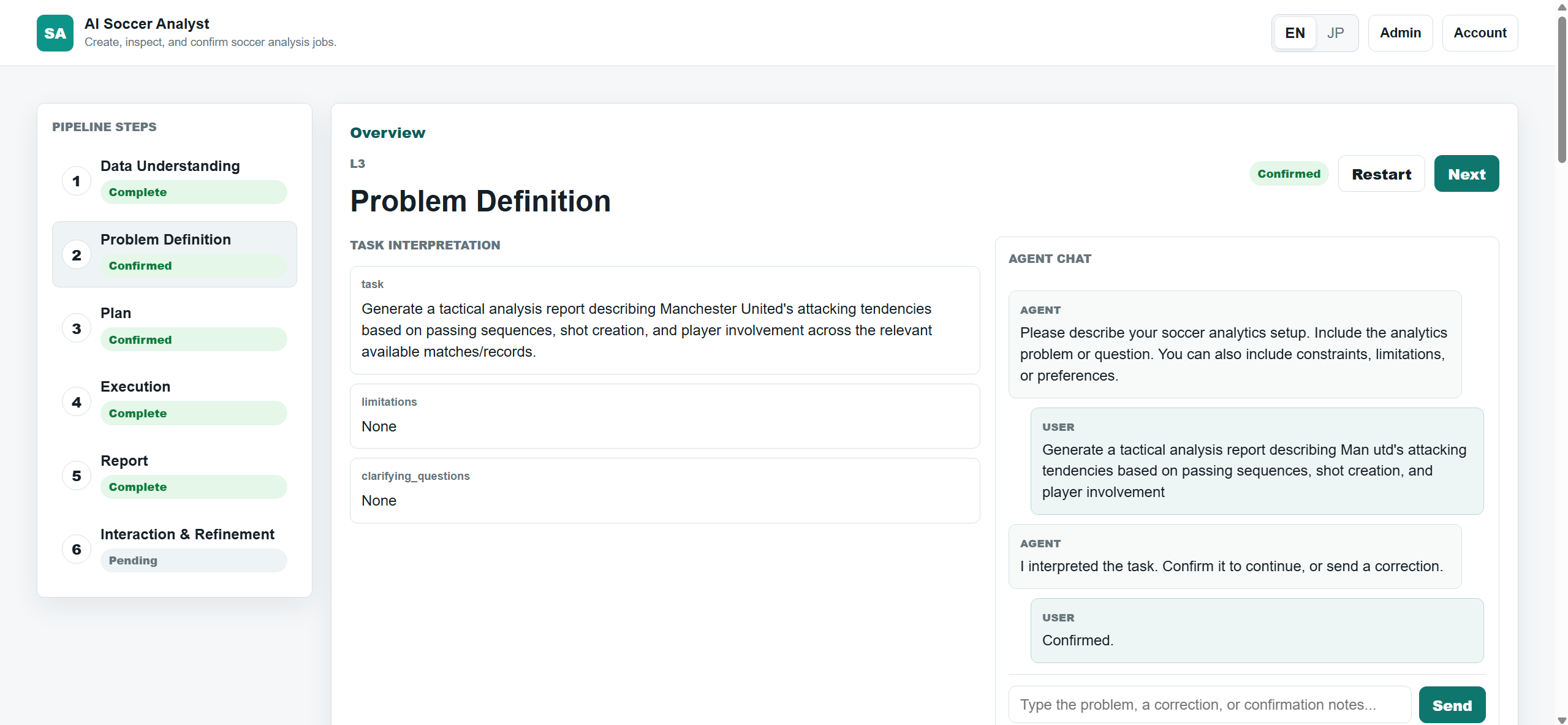}
  \caption{Stage-aware conversation for confirming a problem definition or
  submitting a correction.}
  \label{fig:ui-chat}
\end{figure}

\begin{figure}[!ht]
  \centering
  \includegraphics[width=0.76\linewidth]{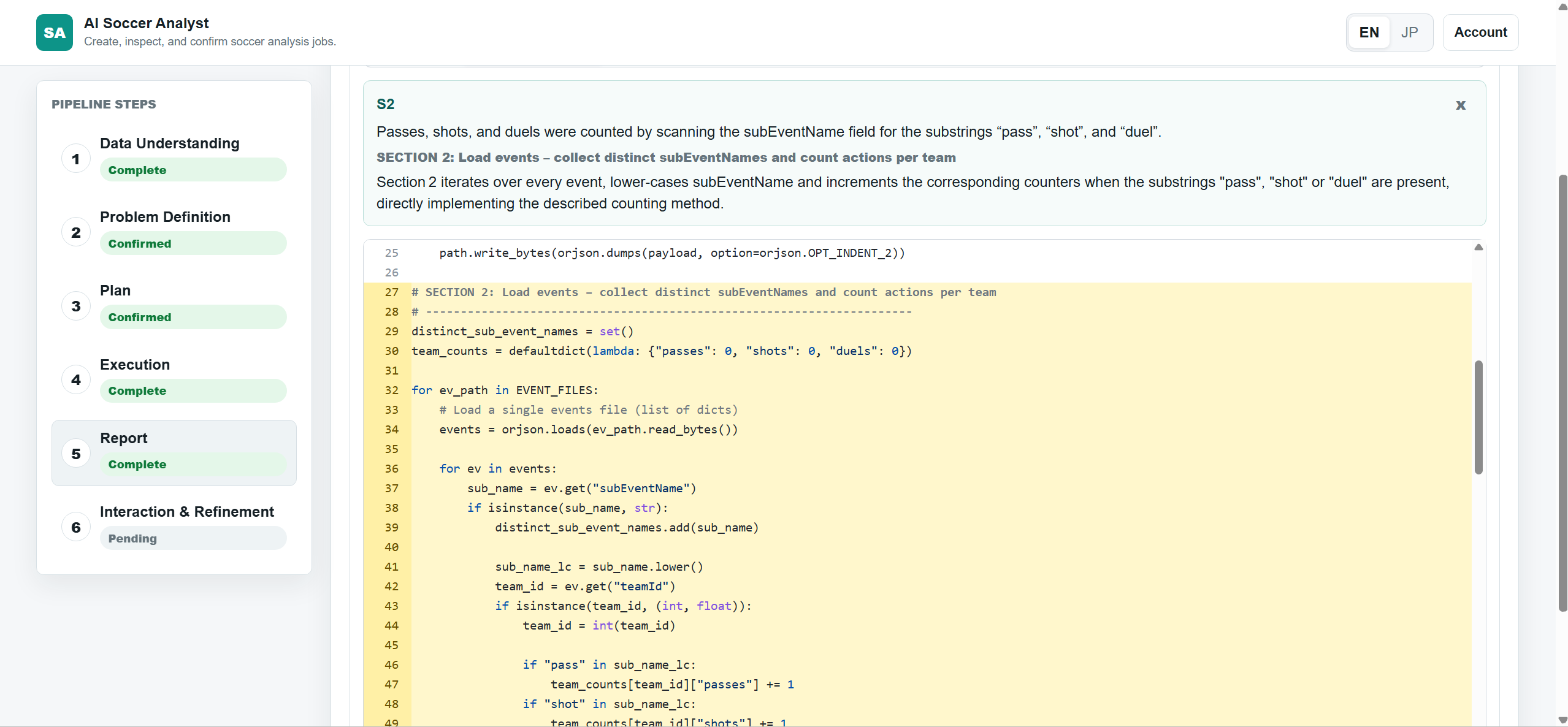}
  \caption{Claim--evidence inspection linking report text to its rationale and
  {related explanatory cleaned-code sections}.}
  \label{fig:ui-evidence}
\end{figure}
{Although Interaction and Refinement introduces no
separate user-facing artifact, the user's request, routing decision,
and confirmation remain visible in the chat history, while the resulting
updated artifact is retained as the current artifact for the responsible
stage.}

\clearpage
\section{Example User-Facing L1 Analysis}
\label{app:l1-example}

This section presents the recorded artifacts for a completed L1
descriptive-retrieval task, including its problem definition, confirmed plan,
report, claim--evidence links, and cleaned-code view. The user's request was to
summarize each team's number of passes, shots, and duels.

\subsection{Job Metadata}
\label{app:l1-metadata}

Table~\ref{tab:l1-metadata} summarizes the recorded configuration and
execution characteristics of this example. ``Complete'' denotes completion of
the automated workflow rather than expert validation of analytical correctness.

\begin{table}[ht]
  \caption{Metadata for the example L1 analysis job.}
  \label{tab:l1-metadata}
  \centering
  \begin{tabularx}{0.92\linewidth}{@{}>{\bfseries}lX@{}}
    \toprule
    Field & Recorded value \\
    \midrule
    Task level & L1 descriptive retrieval \\
    Dataset & Wyscout 2017~\cite{pappalardo2019wyscout} \\
    Final status & Complete (automated workflow) \\
    Elapsed time & 27 minutes 2 seconds \\
    Execution rounds & 9 total: 8 coder revisions and 1 successful completion \\
    Input tokens & 29,405 \\
    Output tokens & 5,772 \\
    \bottomrule
  \end{tabularx}
\end{table}

\subsection{Problem Definition}
\label{app:l1-interpretation}

\uicard[uiuser]{%
  \textcolor{uiaccent}{\bfseries Task}\par\smallskip
  Summarize each team's number of passes, shots, and duels across the relevant
  available matches.
}

\uicard[uiuser]{%
  \textcolor{uiaccent}{\bfseries Limitations}\par\smallskip
  None.
}

\uicard[uiuser]{%
  \textcolor{uiaccent}{\bfseries Clarifying questions}\par\smallskip
  None.
}

\subsection{Confirmed Analysis Plan}
\label{app:l1-plan}

\mdtwo{Goal}

Provide a table that shows, for each team in the dataset, the total number of
passes, shots, and duels performed across all available matches.

\mdtwo{Assumptions}

\begin{itemize}[leftmargin=1.5em]
  \item The events dataset records each action with an \texttt{eventName} field
  that includes values such as \texttt{Pass}, \texttt{Shot}, and \texttt{Duel}
  (or equivalent sub-event names).
  \item Each event row contains a \texttt{teamId} that matches the \texttt{wyId}
  field in the teams metadata file.
  \item All event files under the events folder belong to the matches considered;
  no additional filtering by competition or date is required.
\end{itemize}

\mdtwo{Inputs Needed}

The recorded inputs comprise the seven Wyscout event files for England, France,
Germany, Italy, Spain, the European Championship, and the World Cup, together
with \texttt{teams.json}.

\mdtwo{Steps}

\begin{enumerate}[leftmargin=1.8em]
  \item \textbf{Explore event types.} Load all event files and inspect the
  distinct \texttt{eventName} values, and \texttt{subEventName} if needed, to
  confirm how passes, shots, and duels are recorded.
  \item \textbf{Select relevant actions.} Determine which event or sub-event
  values correspond to passes, shots, and duels.
  \item \textbf{Count actions per team.} Filter the events for the three action
  types and count the rows belonging to each team.
  \item \textbf{Add team names.} Join the counts with the teams metadata so the
  output uses readable team names instead of numeric identifiers.
  \item \textbf{Create the summary table.} Produce columns for team name,
  passes, shots, and duels.
\end{enumerate}

\mdtwo{Expected Output}

\texttt{team\_pass\_shot\_duel\_summary.csv}

\clearpage
\subsection{Generated Report}
\label{app:l1-report}

\mdtwo{Overview}

This report aggregates the total number of \textbf{passes}, \textbf{shots},
and \textbf{duels} performed by every team present in the available Wyscout
2017 event files. Each occurrence of the canonical actions \emph{Pass},
\emph{Shot}, and \emph{Duel} was counted across all event files
\uievidenceid{S3}, and numeric \texttt{teamId} values were mapped to readable
team names from the teams metadata \uievidenceid{S4}.

The final per-team summary is stored in
\texttt{team\_pass\_shot\_duel\_summary.csv}. The excerpt below reproduces the
first 20 rows; the full CSV contains 142 teams.

\begin{center}
\small
\begin{tabular}{@{}lrrr@{}}
\toprule
Team & Passes & Shots & Duels \\
\midrule
Villarreal Club de Fútbol & 16,550 & 435 & 8,519 \\
Levante UD & 12,416 & 352 & 8,456 \\
Real Sociedad de Fútbol & 18,440 & 450 & 8,521 \\
Real Club Celta de Vigo & 18,768 & 386 & 8,515 \\
Getafe Club de Fútbol & 10,692 & 317 & 9,068 \\
Athletic Club Bilbao & 15,022 & 377 & 8,726 \\
Girona FC & 13,508 & 387 & 8,704 \\
Club Atlético de Madrid & 16,336 & 354 & 8,924 \\
Reial Club Deportiu Espanyol & 14,641 & 361 & 8,502 \\
Sevilla FC & 18,027 & 469 & 8,868 \\
SD Eibar & 15,224 & 405 & 9,146 \\
Málaga Club de Fútbol & 13,340 & 344 & 9,016 \\
FC Barcelona & 23,260 & 502 & 7,706 \\
Real Betis Balompié & 19,342 & 359 & 7,919 \\
Real Madrid Club de Fútbol & 22,081 & 631 & 7,860 \\
Real Club Deportivo de La Coruña & 14,539 & 405 & 8,414 \\
Valencia Club de Fútbol & 15,724 & 418 & 8,536 \\
UD Las Palmas & 17,148 & 339 & 8,797 \\
Deportivo Alavés & 11,428 & 331 & 8,729 \\
CD Leganés & 12,236 & 357 & 9,124 \\
\bottomrule
\end{tabular}
\end{center}

\mdtwo{Aggregate Totals}

Across the entire dataset, the actions sum to:
\begin{itemize}[leftmargin=1.5em]
  \item \textbf{Passes:} 1,665,508;
  \item \textbf{Shots:} 43,078; and
  \item \textbf{Duels:} 879,083.
\end{itemize}
These totals result directly from the per-team aggregation
\uievidenceid{S3}.

\mdtwo{Implementation Detail}

The workflow identified the distinct action values \uievidenceid{S1}, mapped
them to the three canonical actions \uievidenceid{S2}, counted actions per
\texttt{teamId} \uievidenceid{S3}, joined readable team names
\uievidenceid{S4}, and exported the final CSV \uievidenceid{S5}.

\mdtwo{Conclusion}

The analysis provides a team-level summary of passes, shots, and duels across
the selected Wyscout 2017 matches. FC Barcelona recorded 23,260 passes, Real
Madrid recorded 22,081 passes, and Real Madrid recorded 631 shots. The complete
per-team summary is available in the accompanying CSV for further inspection.
\clearpage
\subsection{Claim--Evidence Links}
\label{app:l1-evidence}

The evidence view connects report claims to the cleaned-code sections that
explain the corresponding analytical logic. These links expose the
computational basis of each claim but do not independently validate its
analytical interpretation.

\evidencecard{S1}
  {The workflow identified the distinct event and sub-event names.}
  {1: Discover distinct event and sub-event names}
  {The section scans every event file, collects unique \texttt{eventName} and
  \texttt{subEventName} values, and writes the discovered values to JSON.}

\evidencecard{S2}
  {The discovered labels were mapped to the canonical actions Pass, Shot, and
  Duel.}
  {2: Build a mapping to canonical actions}
  {The section first checks for exact action-name matches and uses
  case-insensitive containment only when an exact category is unavailable.}

\evidencecard{S3}
  {Passes, shots, and duels were counted for each team across the event files.}
  {3: Aggregate counts per team}
  {The section constructs a reverse action lookup, iterates through the event
  files, and increments the corresponding per-team action count.}

\evidencecard{S4}
  {Numeric team identifiers were mapped to readable team names.}
  {4: Join counts with team metadata}
  {The section reads \texttt{teams.json}, constructs a
  \texttt{wyId}-to-name mapping, and attaches the selected display name to each
  team's counts.}

\evidencecard{S5}
  {The final summary was exported as a CSV file.}
  {5: Write the CSV summary}
  {The section writes the team name, pass, shot, and duel columns to
  \texttt{team\_pass\_shot\_duel\_summary.csv}.}

\clearpage
\subsection{Cleaned Analysis Code}
\label{app:l1-code}

The executed and cleaned code serve different roles. The executed script
produced the reviewed results. After review, the backend copied that script and
removed pipeline-specific status and control logic while retaining the
analytical processing steps and user-facing output generation. The reporting
stage used the numbered sections below for claim--code linking. Because the
cleaned code was not rerun, it is an explanatory view rather than the
executable provenance record; the reviewed outputs and execution records
remain the evidence for reported claims.

\begin{multicols}{2}
\raggedcolumns
\begin{lstlisting}[
  style=uipython,
  caption={Post-execution cleaned code presented as an explanatory view for the
  L1 team-action summary task.},
  label={lst:l1-clean-code}
]
import orjson
import csv
from pathlib import Path
from collections import defaultdict

# ----------------------------------------------------------------------
# Paths and constants
# ----------------------------------------------------------------------
# Directory where this script resides (job folder)
SCRIPT_DIR = Path(__file__).parent

# Output directory - the current working directory when the script is run
OUTPUT_DIR = Path.cwd()
OUTPUT_DIR.mkdir(parents=True, exist_ok=True)

# Root of the read‑only dataset (as given in the prompt)
DATASET_ROOT = Path("/workspace/data/wyscout2017")


# ----------------------------------------------------------------------
# Helper I/O functions
# ----------------------------------------------------------------------
def write_json(path: Path, payload):
    """Write a JSON payload to *path* using orjson with pretty printing."""
    path.write_bytes(orjson.dumps(payload, option=orjson.OPT_INDENT_2))


def read_json(path: Path):
    """Read a JSON file using orjson."""
    return orjson.loads(path.read_bytes())


# SECTION 1: Discover distinct event and sub‑event names
def discover_event_names():
    """Collect all unique eventName and subEventName values across the dataset."""
    event_names = set()
    sub_event_names = set()

    events_dir = DATASET_ROOT / "events"
    for event_file in events_dir.glob("events_*.json"):
        data = read_json(event_file)
        for rec in data:
            en = rec.get("eventName")
            if isinstance(en, str):
                event_names.add(en)

            sen = rec.get("subEventName")
            if isinstance(sen, str) and sen:
                sub_event_names.add(sen)

    distinct = {
        "eventNames": sorted(event_names),
        "subEventNames": sorted(sub_event_names),
    }

    distinct_path = OUTPUT_DIR / "distinct_event_names.json"
    write_json(distinct_path, distinct)
    return distinct_path, distinct


# SECTION 2: Build a simple mapping to canonical actions (Pass / Shot / Duel)
def build_action_mapping(distinct):
    """
    Create a mapping of canonical action -> list of raw event/sub‑event names.
    The logic first looks for exact matches (case‑sensitive) and, if none are
    found, falls back to case‑insensitive containment.
    """
    mapping = {"Pass": [], "Shot": [], "Duel": []}

    # Helper to add exact matches
    def add_matches(target, candidates):
        for cand in candidates:
            if cand == target:
                mapping[target].append(cand)

    # Exact matches in eventNames
    for target in mapping.keys():
        add_matches(target, distinct["eventNames"])

    # Case‑insensitive containment for any empty categories
    for target, lst in mapping.items():
        if not lst:
            lowered = target.lower()
            for cand in distinct["eventNames"]:
                if lowered in cand.lower():
                    lst.append(cand)
            for cand in distinct["subEventNames"]:
                if lowered in cand.lower():
                    lst.append(cand)

    # Safety net: ensure each action has at least one entry
    for target, lst in mapping.items():
        if not lst:
            lst.append(target)

    mapping_path = OUTPUT_DIR / "action_mapping.json"
    write_json(mapping_path, mapping)
    return mapping_path, mapping


# SECTION 3: Aggregate counts per team for each canonical action
def aggregate_counts(action_mapping):
    """
    Iterate over all event files and count occurrences of each action type
    per teamId. Returns a list of records ready for JSON output.
    """
    # Initialise nested defaultdicts
    team_counts = defaultdict(lambda: {"Pass": 0, "Shot": 0, "Duel": 0})

    # Reverse lookup: raw name -> canonical action
    reverse_lookup = {}
    for action, raw_names in action_mapping.items():
        for raw in raw_names:
            reverse_lookup[raw] = action

    events_dir = DATASET_ROOT / "events"
    for event_file in events_dir.glob("events_*.json"):
        data = read_json(event_file)
        for rec in data:
            team_id = rec.get("teamId")
            if team_id is None:
                continue  # skip malformed rows

            raw_name = rec.get("eventName") or rec.get("subEventName")
            if not raw_name:
                continue

            action = reverse_lookup.get(raw_name)
            if action:
                team_counts[team_id][action] += 1

    # Convert to a list of records for JSON output
    raw_counts = []
    for team_id, counts in team_counts.items():
        raw_counts.append({
            "teamId": team_id,
            "passes": counts["Pass"],
            "shots": counts["Shot"],
            "duels": counts["Duel"],
        })

    raw_path = OUTPUT_DIR / "team_action_counts_raw.json"
    write_json(raw_path, raw_counts)
    return raw_path, raw_counts


# SECTION 4: Join aggregated counts with team metadata
def join_team_names(raw_counts):
    """Replace team IDs with human‑readable team names."""
    teams_path = DATASET_ROOT / "teams.json"
    teams_data = read_json(teams_path)

    # Build a mapping wyId -> display name (prefer officialName, fallback to name)
    team_name_map = {}
    for team in teams_data:
        wy_id = team.get("wyId")
        name = team.get("officialName") or team.get("name") or f"team_{wy_id}"
        team_name_map[wy_id] = name

    summary = []
    for rec in raw_counts:
        team_id = rec["teamId"]
        team_name = team_name_map.get(team_id, f"team_{team_id}")
        summary.append({
            "team_name": team_name,
            "passes": rec["passes"],
            "shots": rec["shots"],
            "duels": rec["duels"],
        })

    summary_path = OUTPUT_DIR / "team_action_summary.json"
    write_json(summary_path, summary)
    return summary_path, summary


# SECTION 5: Write CSV summary for end‑users
def write_csv(summary):
    """Create a CSV file containing the final per‑team action summary."""
    csv_path = OUTPUT_DIR / "team_pass_shot_duel_summary.csv"
    with csv_path.open("w", newline="", encoding="utf-8") as f:
        writer = csv.DictWriter(f, fieldnames=["team_name", "passes", "shots", "duels"])
        writer.writeheader()
        for row in summary:
            writer.writerow({
                "team_name": row["team_name"],
                "passes": row["passes"],
                "shots": row["shots"],
                "duels": row["duels"],
            })
    return csv_path


# ----------------------------------------------------------------------
# Main orchestration (cleaned - no execution‑pipeline reporting)
# ----------------------------------------------------------------------
def main():
    # SECTION 1
    distinct_path, distinct = discover_event_names()

    # SECTION 2
    mapping_path, action_mapping = build_action_mapping(distinct)

    # SECTION 3
    raw_counts_path, raw_counts = aggregate_counts(action_mapping)

    # SECTION 4
    summary_path, summary = join_team_names(raw_counts)

    # SECTION 5
    write_csv(summary)


if __name__ == "__main__":
    main()
\end{lstlisting}
\end{multicols}

\end{document}